\documentclass[10pt,conference]{IEEEtran}
\AtBeginDocument{%
  \providecommand\BibTeX{{%
    \normalfont B\kern-0.5em{\scshape i\kern-0.25em b}\kern-0.8em\TeX}}}

\usepackage{tabularx}
\usepackage{booktabs}
\usepackage{varwidth}

\usepackage{xspace}

\usepackage{enumitem}

\usepackage{listings}
\usepackage[T1]{fontenc}
\usepackage{upquote}
\usepackage[scaled]{beramono}
\usepackage{xcolor} % For color support

\usepackage{xcolor}
\definecolor{mygreen}{RGB}{0,128,0}

\lstdefinelanguage{diff}{
  basicstyle=\ttfamily\footnotesize,
  morecomment=[f][\color{mygreen}]{+}, % Added lines
  morecomment=[f][\color{red}]{-}, % Removed lines
  morecomment=[f][\color{gray}]{@@}, % Hunk delimiters
}

\definecolor{ToolColor}{rgb}{0.0, 0.2, 0.6}
\colorlet{FixerColor}{mygreen}
\definecolor{InferenceColor}{rgb}{0.6, 0.1, 0.1}
\usepackage{titlesec}
\titleformat{\subsubsection}[runin]
  {\normalfont\normalsize\bfseries}{\thesubsubsection}{.3em}{}
\titlespacing*{\subsubsection}{0pt}{0pt}{.5em}

\usepackage{tabularx} % Add to preamble
\usepackage{subcaption}

\usepackage{tikz}
\usepackage{setspace,caption}
\usepackage{flushend} % balanced columns on last page
\usepackage{hyperref}
\usepackage{url} % URLs; used in plume-bib

\usepackage{amsmath}
\usepackage{amssymb}
\usepackage{mathtools}
\usepackage{amsfonts}
\usepackage{booktabs}
\usepackage{multirow}
\usepackage{siunitx} 
\usepackage{graphicx}

\usepackage{cleveref}

\usepackage{tikz}
\usetikzlibrary{positioning}
\usetikzlibrary{calc}

\usepackage{relsize}

\usepackage{amsthm}

\usepackage[final]{microtype}

\newtheorem{definition}{Definition}[subsubsection]
\begin{document}

%% Alternate formulation; use whichever of the two you prefer:
% \newcommand{\tool}{the Type Inference Framework\xspace}
% \newcommand{\Tool}{The Type Inference Framework\xspace}
% \newcommand{\toolShort}{TIF\xspace}

\newcommand{\conferencePageLimit}{11}

%%% Todo comments
\newcommand{\todo}[1]{{\color{red}\bfseries [[#1]]}}
%% Comment or uncomment this line.
\renewcommand{\todo}[1]{\relax}
% Don't show todo commands if the \notodocomments macro is defined.
\ifdefined\notodocomments
  \renewcommand{\todo}[1]{\relax}
\fi

\newcommand{\manu}[1]{\todo{Manu: #1}}
\newcommand{\mde}[1]{\todo{MDE: #1}}

% Include text in \iflongversion...fi if the \createlongversion macro is defined.
\newif\iflongversion
\ifdefined\createlongversion
  \longversiontrue
\fi
\newcommand{\longversion}[1]{\iflongversion #1\else\relax\fi}

% Use like: \ifanonymous{ANONYMOUS TEXT}\else{NON-ANONNYMOUS TEXT}\fi
% where the "\else{NON-ANONNYMOUS TEXT}" may be omitted.
\newif\ifanonymous
%% Comment or uncomment this line
\anonymoustrue

\newcommand{\anonurl}[1]{\ifanonymous URL removed for anonymity.\else\url{#1}\fi}
\newcommand{\footnoteanonurl}[1]{\footnote{\anonurl{#1}}}

% \|name| or \mathid{name} denotes identifiers and slots in formulas
\def\|#1|{\mathid{#1}}
\newcommand{\mathid}[1]{\ensuremath{\mathit{#1}}}
% \<name> or \codeid{name} denotes computer code identifiers
\def\<#1>{\codeid{#1}}
% \protected\def\codeid#1{\ifmmode{\mbox{\sf{#1}}}\else{\sf #1}\fi}
% \protected\def\codeid#1{\ifmmode{\mbox{\ttfamily{#1}}}\else{\ttfamily #1}\fi}
\protected\def\codeid#1{\ifmmode{\mbox{\smaller\ttfamily{#1}}}\else{\smaller\ttfamily #1}\fi}
% \protected\def\codeid#1{\mintinline{java}{#1}}

% research question list, based on the answer to https://tex.stackexchange.com/questions/559305/how-to-format-for-two-column-research-question
\newlist{researchquestions}{enumerate}{1}
\setlist[researchquestions]{label*=\textbf{RQ\arabic*}, leftmargin=*}

\newcommand{\CalledMethodsBottom}{\<@Call\-ed\-Meth\-ods\-Bottom>\xspace}
\newcommand{\CalledMethods}{\<@Call\-ed\-Meth\-ods>\xspace}
\newcommand{\EnsuresCalledMethods}{\<@En\-sures\-Call\-ed\-Meth\-ods>\xspace}
\newcommand{\MustCall}{\codeid{@Must\-Call}\xspace}
\newcommand{\MustCallAlias}{\codeid{@Must\-Call\-Alias}\xspace}
\newcommand{\MustCallUnknown}{\codeid{@Must\-Call\-Unknown}\xspace}
\newcommand{\CreatesMustCallFor}{\<@Creates\-Must\-Call\-For>\xspace}
\newcommand{\Owning}{\<@Own\-ing>\xspace}
\newcommand{\NotOwning}{\<@Not\-Own\-ing>\xspace}
% Deprecated
\newcommand{\ResetMustCall}{\CreatesMustCallFor}

% "trule" stands for ``type rule''
\newcommand{\trule}[2]{\[\frac{#1}{#2}\]}
\newcommand{\truleinline}[2]{\ensuremath{#1\mathrel{\vdash}#2}}
\newcommand{\hastype}[1]{\mathbin{:}\trtext{#1}}
\newcommand{\trcode}[1]{\codeid{\smaller\smaller #1}}
\newcommand{\trtext}[1]{\mbox{\smaller\smaller #1}}
\newcommand{\trquoted}[1]{\trcode{"}#1\trcode{"}}

%%% Computed values

% For any number that's referenced in the text itself (and a table), create a macro like these rather than copy-pasting.
% These examples are from the WPI paper; you can delete them.

\newcommand{\numTypeSystems}{11\xspace} % Formatter, index, interning, lock, nullness, regex, resourceleak, signature, signedness, InitializedFields, Optional
\newcommand{\numModifiedTypeSystems}{2\xspace} % Formatter, Nullness (not Called Methods, because the postcondition code is general)
\newcommand{\numProjects}{12\xspace}
\newcommand{\numLOC}{88,680\xspace}
\newcommand{\numHumanAnnos}{803\xspace}
\newcommand{\percentInferred}{39\todo{check}\%\xspace}
\newcommand{\warningReductionPercent}{45\todo{check}\%\xspace}
\newcommand{\tsSpecificLoC}{61\todo{check}\xspace}

%%% Miscellaneous

\hyphenation{type-state}        % LaTeX defaults to "types-tate"
\hyphenation{null-able}         % LaTeX defaults to "nul-lable"
\hyphenation{Spot-Bugs}

%%% Space-saving hacks

% Reduce indentation in lists.
\setlength{\leftmargini}{.75\leftmargini}
\setlength{\leftmarginii}{.75\leftmarginii}
\setlength{\leftmarginiii}{.75\leftmarginiii}

\newcommand{\prefigcaption}{\vspace{-5pt}}
\newcommand{\posttablecaption}{\vspace{-5pt}}

% Reduce the separation between figures and text.
\addtolength{\textfloatsep}{-.25\textfloatsep}
\addtolength{\dbltextfloatsep}{-.25\dbltextfloatsep}
\addtolength{\floatsep}{-.25\floatsep}
\addtolength{\dblfloatsep}{-.25\dblfloatsep}

\newcommand{\zph}{\phantom{0}}
\newcommand{\zzph}{\phantom{00}}

\newcommand{\ie}{i.e.,\xspace}
\newcommand{\eg}{e.g.,\xspace}

\newcommand{\GenerateDeterminantFromMinor}{\<Gener\-ate\-De\-ter\-mi\-nant\-From\-Minor>\xspace}
\newcommand{\PrintStream}{\<Print\-Stream>\xspace}
\newcommand{\PrintWriter}{\<Print\-Writer>\xspace}
\newcommand{\trycatch}{\<try>--\<catch>\xspace}
\newcommand{\tryfinally}{\<try>--\<finally>\xspace}

\newcommand{\OurTool}{Arodnap\xspace}
\newcommand{\ourTool}{Arodnap\xspace}

\newcommand{\RLCRLFixer}{RLC+\discretionary{}{}{}RLFixer\xspace}
\newcommand{\RLCIRLCRLFixer}{RLCI+\discretionary{}{}{}RLC+\discretionary{}{}{}RLFixer\xspace}

\newcommand{\RCA}{Field Containment Analysis\xspace} % Only for section headers
\newcommand{\rca}{field containment analysis\xspace}
\newcommand{\Rca}{Field containment analysis\xspace}
\newcommand{\rc}{field containment\xspace}
\newcommand{\Rc}{Field containment\xspace}

% RLC+RLFixer
\newcommand{\BOneCL}{1909}
\newcommand{\BOneXE}{0}
\newcommand{\BOneXR}{0}
\newcommand{\BOneTot}{1909}
\newcommand{\BOneFixCL}{783}
\newcommand{\BOneFixXE}{0}
\newcommand{\BOneFixTot}{783}
\newcommand{\BOneR}{41}

% RLCI+RLC+RLFixer
\newcommand{\BTwoCL}{1537}
\newcommand{\BTwoXE}{320}
\newcommand{\BTwoXR}{356}
\newcommand{\BTwoTot}{2213}
\newcommand{\BTwoFixCL}{755}
\newcommand{\BTwoFixXE}{5}
\newcommand{\BTwoFixTot}{760}
\newcommand{\BTwoR}{50}

% \OurTool
\newcommand{\BestCL}{1446}
\newcommand{\BestXE}{243}
\newcommand{\BestXR}{447}
\newcommand{\BestTot}{2136}
\newcommand{\BestFixCL}{952}
\newcommand{\BestFixXE}{62}
\newcommand{\BestFixTot}{1014}
\newcommand{\BestR}{68}

% Ablation 1 (-- Code Transformations)
\newcommand{\AOneCL}{1537}
\newcommand{\AOneXE}{253}
\newcommand{\AOneXR}{361}
\newcommand{\AOneTot}{2151}
\newcommand{\AOneFixCL}{951}
\newcommand{\AOneFixXE}{54}
\newcommand{\AOneFixTot}{1005}
\newcommand{\AOneR}{64}

% Ablation 2 (-- RLFixer Enhancements)
\newcommand{\ATwoCL}{1477}
\newcommand{\ATwoXE}{253}
\newcommand{\ATwoXR}{418}
\newcommand{\ATwoTot}{2148}
\newcommand{\ATwoFixCL}{755}
\newcommand{\ATwoFixXE}{44}
\newcommand{\ATwoFixTot}{799}
\newcommand{\ATwoR}{57}

% Ablation 3 (-- Field Overwrite Handling)
\newcommand{\AThreeCL}{1478}
\newcommand{\AThreeXE}{324}
\newcommand{\AThreeXR}{414}
\newcommand{\AThreeTot}{2216}
\newcommand{\AThreeFixCL}{967}
\newcommand{\AThreeFixXE}{5}
\newcommand{\AThreeFixTot}{972}
\newcommand{\AThreeR}{63}

% At least 90% of every float page must be taken up by
% floats; there will be no page with more than 10% white space.
\def\topfraction{.9}
\def\dbltopfraction{\topfraction}
\def\floatpagefraction{\topfraction}     % default .5
\def\dblfloatpagefraction{\topfraction}  % default .5
\def\textfraction{.1}

\newcommand{\revised}[1]{{\color{blue}#1}} % using a different color to differentiate from the TODOs
\renewcommand{\revised}[1]{#1}
% Left and right curly braces in tt font
\newcommand{\ttlcb}{\texttt{\char "7B}}
\newcommand{\ttrcb}{\texttt{\char "7D}}

% LocalWords:  researchquestions RQ leftmargin

\title{Repairing Leaks in Resource Wrappers}

\author{%
\IEEEauthorblockN{Sanjay Malakar\IEEEauthorrefmark{1},
Michael D. Ernst\IEEEauthorrefmark{2},
Martin Kellogg\IEEEauthorrefmark{3},
Manu Sridharan\IEEEauthorrefmark{1}}
\IEEEauthorblockA{\IEEEauthorrefmark{1}University of California, Riverside, USA \quad
\IEEEauthorrefmark{2}University of Washington, USA \quad
\IEEEauthorrefmark{3}New Jersey Institute of Technology, USA}
% Use the next email list to merge Sanjay's and Manu's emails into one block
% \IEEEauthorblockA{\small \textit{Email:}
% \texttt{\{smala009, manus\}@ucr.edu},
% \texttt{martin.kellogg@njit.edu},
% \texttt{mernst@cs.washington.edu}}
\IEEEauthorblockA{\small \textit{Email:}
\texttt{smala009@ucr.edu},
\texttt{mernst@cs.washington.edu},
\texttt{martin.kellogg@njit.edu},
\texttt{manu@cs.ucr.edu}}
}

\maketitle


\begin{abstract}
%Resource leaks are a common cause of software failures.
A resource leak occurs when a program fails to release a finite
resource like a socket, file descriptor or database
connection.
While sound static analysis tools
can \emph{detect} all leaks,
automatically \emph{repairing} them remains challenging.
Prior work took the output of a detection tool and
attempted to repair only leaks from a hard-coded list of library resource
types.
% that are known to the repair tool \emph{a priori}.
That approach limits the scope of repairable leaks:
real-world code uses \emph{resource wrappers} that store a resource in a field
and must themselves be closed.

This paper makes four key contributions to improve resource leak repair in the presence of wrappers.
(1) It integrates inference of resource management specifications into the repair pipeline, enabling extant fixing approaches to reason about wrappers.
(2) It transforms programs into variants that are easier to analyze, making
inference, detection, and fixing tools more effective; for instance,
it makes detection tools report problems closer to the root cause,
often in a client of a resource wrapper rather than within the wrapper class itself.
(3) A novel \emph{\rca} reasons about resource lifetimes,
enabling repair of more leaks involving resources stored in fields.
(4) It introduces a new repair pattern and more precise reasoning to better handle resources stored in non-final fields.

Prior work fixed \BOneR\% of resource leak warnings in the NJR benchmark suite;
our implementation \ourTool fixes \BestR\%.

\label{abstract:dummy-for-etags}

% LocalWords:  NJR

\end{abstract}

\maketitle

% \mde{The paper is inconsistent about using "RLC" or "the RLC" as a noun.  I think the former flows better, but in any event make the paper consistent.  If we do choose "the RLC", we should probably use "the RLCI" for consistency.}\manu{I think we should go with ``RLC''; I will edit accordingly}

\section{Introduction}
\label{sec:intro}
Resource leaks, such as unreleased file handles, sockets, or database connections, are a persistent source of reliability issues.
These defects often evade detection during testing and manifest only after extended run time, leading to degraded performance, outages, and even security
vulnerabilities~\cite{ZhangUSP2024, ChristakisB2016}. Static analysis is a powerful tool for detecting such defects early in development, and many modern
analyzers for Java\,---\,such as Infer~\cite{CalcagnoDDGHLOPPR2015}, SpotBugs~\cite{HovemeyerP2004}, and the Checker Framework's Resource Leak Checker~\cite{KelloggSSE2021}\,---\,can report
potential leaks based on ownership and control-flow reasoning.

However, detection alone is insufficient. Developers complain that static analyses surface true issues but fail to provide actionable
suggestions~\cite{ChristakisB2016} and have too many false positives \cite{JohnsonSMHB2013}. Automated repair tools
that rely only on test-based validation \cite{LeGouesNFW2012,vanTonderLG2018}
are ineffective for leak repair, because resource management leaks typically do not manifest in test failures.
Many leaks share common manifestation and repair
patterns~\cite{GhanavatiKAP2020}; the state-of-the-art tool
RLFixer~\cite{UttureP2023} takes advantage of this insight. RLFixer relies on extant leak-detection tools
like Infer, SpotBugs, and the Resource Leak Checker to detect leaks, which
it repairs using a fixed set of repair templates.
However, because RLFixer treats the leak-detection tools as a black-box warning oracle,
it is limited to repairing leaks of \emph{library resources}, like sockets or file descriptors,
that the leak-detection tool tracks by default.

Our key insight is that repairing only leaks of library resources is
usually insufficient, because many resources in practice are managed by
\emph{wrappers}:
programmer-written classes that themselves act as resources, like the examples
in \cref{lst:inference-example,fig:mot-example}.
Acting on this insight requires detecting and reasoning about such wrappers during the leak-detection stage,
but extant leak-detection tools only reason about library resources (e.g., those
defined in the JDK) by default.
% {\,---\,}they require programmers to write specifications
% for their wrappers to enable reasoning about them.
%
Recent work has extended one detection tool{\,---\,}the Resource Leak
Checker (RLC){\,---\,}with a specification
generator (``RLC Inference'' or ``RLCI'') that automates this manual process~\cite{ShadabGTEKLLS2023}.
Our first contribution is to extend the combination of RLFixer and RLC
to take advantage of RLCI specifications.
This makes it possible to automatically detect and fix leaked programmer-written wrappers instead of only
library resources.

However, the \RLCIRLCRLFixer combination is only marginally better
than the base \RLCRLFixer combination, repairing \BTwoR\% instead of \BOneR\% of leaks
in our experiments.  The reason is that many resource leaks involving wrappers require
reasoning about resources stored in \emph{fields}.
RLFixer marks a leak as ``unfixable'' whenever a resource
may escape into a field.
Our remaining contributions are a set of new program transformations and analyses,
embodied in a tool called \ourTool. \OurTool extends RLFixer to reason about resources
that are stored in fields, as well as fix leaks resulting from mishandling of those
resources.

\OurTool improves on the handling of fields in prior work in three main ways.
(1) It adds a code transformation stage to the detection-and-fixing
pipeline (\cref{fig:overview-pipeline})
that runs after inference and leak detection.
This code transformation stage
enables precise reasoning and inference for many 
resource-containing fields; for example, it adds the \<final> qualifier
to eligible fields, converts resource-containing fields that 
\revised{can be scoped to a single method}
% are only used in one method
into local variables, and adds missing
finalizer methods to classes that contain a resource field.
After these transformations, inference and leak detection
generate more actionable leak warnings.
(2) We enhanced RLFixer to reason about fields: a new \emph{\rca}
makes RLFixer's escape logic sound and less
conservative in its field handling.
(3) We improved handling of field overwrites.
For overwrites in constructors, we improved RLC's reasoning.
For overwrites in other methods, we added a new repair
pattern to RLFixer, along with an analysis to
determine when the new repair can be applied soundly.
With these improvements, \ourTool can resolve \BestR\% of all
resource leak warnings on the same set of benchmark programs used in RLFixer's
evaluation, vs. \BOneR\% for RLFixer alone.
In sum, our contributions are:
\begin{itemize}
  \item We extend RLFixer to use RLCI to reason about wrappers.
  \item We introduce code transformations to ease analysis of and inference
    for resource fields.
  \item We introduce \emph{\rca}, a static analysis that identifies
    resource wrapper classes whose internal fields do not escape, enabling
    sound repair of more leaks.
  \item We improve analysis and repair for leaks due to non-final resource fields.
  \item We implemented our approach in a tool \ourTool and evaluated it on
    the NJR dataset~\cite{palsberg2018njr} originally used to evaluate RLFixer,
    improving the fix rate from \BOneR\% for RLFixer to \BestR\% for \ourTool.
\end{itemize}
Our artifact includes all the code, data, and scripts used in this paper~\cite{artifact}.
\OurTool itself is also open-source~\cite{arodnap}.

\section{Background}
This section describes the three state-of-the-art tools on which our work builds:
RLC for detecting resource leaks (\S\ref{sec:rlc}),
RLCI for inferring resource specifications (\S\ref{sec:inference}),
and RLFixer (\S\ref{sec:rlfixer}) for suggesting repairs.
It explains how these tools operate,
their limitations, and the motivation for the enhancements introduced in this work.

\subsection{Static Leak Detection with the Resource Leak Checker}
\label{sec:rlc}
The Resource Leak Checker (RLC) is a pluggable type system built on top of the Checker Framework~\cite{PapiACPE2008, KelloggSSE2021}. It verifies that objects
such as files, sockets, or streams are cleaned up (by an explicit call to a
finalizer method) before they become unreachable. RLC is sound by design,
scales to real-world codebases, and requires a manageable annotation burden.
% \mde{word choice}
%

At the core of RLC is the notion of \MustCall obligations. A type \<T> annotated with \MustCall\<("close")> indicates that \<close()>
must be invoked on every \<T> object before its lifetime ends.
The RLC analysis ensures that all such required methods are called along every path that leads to the object
becoming unreachable (e.g., via scope exit or variable overwrite).
Programmer-written resource specifications (expressed as RLC annotations)
express ownership, obligation transfer, and aliasing relationships.
\begin{itemize}
    \item \Owning references are responsible for eventually satisfying the \MustCall obligations of the object they refer to.
    \item \NotOwning marks a reference that is not responsible for the obligation, such as shared or borrowed values.
%     \item \MustCallAlias indicates that multiple references of different
%       Java types have a single underlying obligation.  Calling the required method on \emph{any} one of
%     them is sufficient to satisfy the obligation.\mde{Specifically mention
%       resource wrappers here, since that is the focus of the paper and is
%       in the title.}\mde{There is no further mention of \MustCallAlias in
%       the paper.  How about dropping it here, for brevity and to avoid
%       distracting the reader?}
%       \todo{SM: Does mentioning resource wrappers only here make sense? As this bullet is about MustCallAlias ony. Wrappers could also be detected from Owning and EnsuresCalledMethods}
    \item \EnsuresCalledMethods\<(x,> \<y)>, written on a method $m$,
      guarantees that \<x.y()> is called before $m$ returns.
\end{itemize}
\begin{figure}
\begin{lstlisting}[language=diff]
+@MustCall("close")
 class MyWriter {
+  @Owning PrintWriter pw;
   MyWriter(String path) {
     pw = new PrintWriter(path);
   }
+  @EnsuresCalledMethods(value="pw", methods="close")
   void close() {
     pw.close();
   }
 }
 void use() {
   MyWriter writer = new MyWriter("f.txt");
 }
\end{lstlisting}
\caption{Specification inference makes leaks repairable.
  With no resource management specification annotations, RLC reports a leak at line~5,
  where repair is not possible.
  With added specifications (highlighted in \textcolor{FixerColor}{green}),
  RLC reports the leak at line 13{\,---\,}where repair is
  feasible by closing the \<MyWriter> object.}
\label{lst:inference-example}
\end{figure}

\label{fig:inference-example:dummy-for-etags}

% LocalWords:  lst MyWriter PrintWriter pw

%
These annotations allow RLC to track resource lifecycle responsibilities
across field assignments, parameter passing, method returns, and resource wrappers. Rather than relying
on whole-program alias analysis, RLC uses these annotations to reason about ownership and aliasing in a modular, sound way~\cite{KelloggSSE2022}.
To statically verify that obligations are fulfilled, RLC runs three cooperating analyses~\cite{KelloggSSE2021}. First, a type system computes the set
of required \MustCall methods for each reference. Second, a type system
computes which methods have definitely been invoked on a given value. Finally, a dataflow analysis verifies that all required methods have been called before a resource becomes unreachable.

RLC is a specify-and-verify system:  the programmer writes explicit
ownership and aliasing annotations, and the programmer obtains a sound
guarantee of no resource leaks.  Unfortunately, most Java codebases lack
these specifications. RLC ships with specifications for
the JDK standard library; programmers are
expected to write annotations like \Owning and \MustCallAlias to specify
their own resource-handling code.
Writing these annotations is tedious and error-prone, motivating
automatic inference of resource-management specifications.

\lstset{
  escapeinside={(*@}{@*)},      % anything between (*@ ... @*) is parsed as LaTeX
}

% handy macros for code-colouring
\newcommand{\cthl}[1]{\textcolor{ToolColor}{#1}}
\newcommand{\fixhl}[1]{\textcolor{FixerColor}{#1}}
\newcommand{\infhl}[1]{\textcolor{InferenceColor}{#1}}

\begin{figure*}[!t]
\centering

\begin{subfigure}[t]{0.48\textwidth}
\begin{lstlisting}[language=Java]
class TempFileWriter {
  private PrintStream stream;

  public TempFileWriter(String path) {
      stream = new PrintStream(path);
  }

  void resetStream(String path) {

    stream = new PrintStream(path);
  }

  public void printSomething() {
    stream.println("hello");
  }




}

class Client {
  public static void print() {


    TempFileWriter tmp = new TempFileWriter("f.txt");
    tmp.printSomething();



  }
}
\end{lstlisting}
\caption{Leaky version without cleanup. RLC reports leaks at lines 5 and 10.}
\label{lst:mot-example-leaky}
\end{subfigure}
\hfill
\begin{subfigure}[t]{0.48\textwidth}
\begin{lstlisting}[language=Java]
class TempFileWriter (*@\cthl{implements AutoCloseable}@*) {
  (*@\infhl{@Owning}@*) private PrintStream stream;

  public TempFileWriter(String path) {
      stream = new PrintStream(path);
  }

  void resetStream(String path) {
    (*@\fixhl{if (stream != null) stream.close();}@*)
    stream = new PrintStream(path);
  }

  public void printSomething() {
    stream.println("hello");
  }

  (*@\cthl{public void close() \{ }@*)
    (*@\cthl{stream.close();}@*)
  (*@\cthl{\}}@*)
}

class Client {
  public static void print() {
    TempFileWriter tmp = null;
    (*@\fixhl{try \{}@*)
      tmp = new TempFileWriter("f.txt"); (*@\label{line:newTempFileWriter}@*)
      tmp.printSomething();
    (*@\fixhl{\} finally \{}@*)
      (*@\fixhl{if (tmp != null) tmp.close();}@*)
    (*@\fixhl{\}}@*)
  }
}
\end{lstlisting}
\caption{\OurTool
  inserts a finalizer (in \textcolor{ToolColor}{blue}),
  letting inference infer ownership (in \textcolor{InferenceColor}{red}), 
  shifting RLC's leak warning from line 5 to line \ref{line:newTempFileWriter},
  enabling RLFixer to insert cleanup logic
(in \textcolor{FixerColor}{green}).}
\label{lst:mot-example-fixed}
\end{subfigure}

\caption{Simplified leak repair example from the NJR dataset,
benchmark \<url882f91ec97\_WenboCao\_Microsoft\_Drone>, file
\<Generate\-DeterminantFromMinor.java>.
RLFixer cannot repair
the original code (\cref{lst:mot-example-leaky}).
In \cref{lst:mot-example-fixed} \OurTool inserts the missing finalizer,
which enables inference to infer ownership.
RLFixer then eliminates both leaks:  one inserting a \tryfinally wrapper at
lines 25--30
and one by inserting a \emph{pre-close} on line 9 before the field overwrite.
}
% \mde{The paper is inconsistent about what pre-close looks like.  Here there
  % is no \<try>--\<catch>, but in \cref{lst:mot-example-fixed} there is. SM: The pre-close has catch in the actual fix, but I simplified it here to save space. I'll do the same for the other example.}

\label{fig:mot-example}
\end{figure*}

% LocalWords:  escapeinside TempFileWriter resetStream printSomething tmp
% LocalWords:  println NJR RLC's url882f91ec97 WenboCao
% LocalWords:  DeterminantFromMinor

\subsection{Inference for Resource Management Annotations}
\label{sec:inference}

RLC Inference (RLCI)~\cite{ShadabGTEKLLS2023} statically discovers
specifications related to resource lifecycles and ownership. By analyzing
how objects are allocated, passed, and used throughout a program, the
inference identifies and automatically annotates patterns of
resource management.

This approach is especially valuable when analyzing a user-defined resource
\emph{wrapper}\,---\,a class that internally manages a resource delegate but may not explicitly expose it.
In such cases, inference can insert annotations documenting that 
the wrapper owns its internal resources and exposes a finalizer method like
\<close()> that satisfies their obligations. These inferred specifications
enable RLC to verify both the implementation and the use sites of the
wrapper class.

\Cref{lst:inference-example} demonstrates how inference enables more ac\-tion\-able leak warnings. The class \<My\-Writer> internally allocates a \PrintWriter and provides
a \<close()> method that properly closes it. Without inference, RLC
does not recognize that \<My\-Writer> is a wrapper that is
responsible for closing the resource.  RLC therefore reports a leak at the \PrintWriter
allocation inside the constructor (because the \PrintWriter is not closed
before the constructor returns). Once inference recovers the necessary specifications (in green in \cref{lst:inference-example}), including \MustCall on the class and \EnsuresCalledMethods on the cleanup method, RLC shifts
the warning to the \<use()> method, which erroneously allocates a \<My\-Writer> object without closing it.

By \emph{shifting} the warning from inside the wrapper to its call site, inference makes the leak warning more actionable and amenable to automatic repair.

\subsection{Semi-Automated Repair with RLFixer}
\label{sec:rlfixer}
RLFixer~\cite{UttureP2023} automatically generates fix hints for resource leaks
based on warnings emitted by static analysis tools, such as RLC or Infer~\cite{CalcagnoDDGHLOPPR2015}, which it treats as black-box warning generators.
It aims to
% \mde{Does it succeed?  Be explicit.}
% generate semantically sound, minimally invasive
generate repairs that do not alter core program logic, using
control-flow scaffolding like \tryfinally blocks to enforce cleanup.
RLFixer emits its output as textual hints without applying or validating them, requiring manual developer intervention to write patches.

For each leak warning, RLFixer performs a lightweight alias analysis to
group variables referring to the same resource instance{\,---\,}handling
both direct aliases and a special case of
user-defined wrapper types (those that encapsulate a resource via constructor
injection and a \<close()> method).
It then conducts a demand-driven resource escape analysis \cite[\S3.3]{UttureP2023} to
determine whether the resource escapes its enclosing method through fields, return values, method parameters, or data structures. If any such escape is detected, RLFixer conservatively marks
the leak as unrepairable; %, since a safe cleanup cannot be guaranteed within the local context.
inserting a \<close()> call at the warning site in such cases risks a ``use-after-close'' error, because external code might still use the resource afterward.

After deeming a leak fixable, RLFixer selects a repair template from a set
of hard-coded options. Each involves inserting a \<close()> call in a \<finally> block, optionally wrapping
existing resource usage in a structured control-flow construct such as a
\trycatch.

% \manu{Do we need the next sentence?  Later we say double-close is usually
% fine, so this is slightly confusing}RLFixer removes redundant \<close()>
% calls within the same scope to avoid a double-close or unintended behaviors.

\usetikzlibrary{arrows.meta, positioning, fit, backgrounds}
\pgfdeclarelayer{foreground}
\pgfsetlayers{background,main,foreground}

\definecolor{newcolor}{RGB}{232,255,232}
\definecolor{enhancedcolor}{RGB}{255,255,232}

\definecolor{lightblue}{RGB}{232,244,248}
\definecolor{lightyellow}{RGB}{255,255,232}
\definecolor{lightgreen}{RGB}{232,255,232}
\definecolor{lightred}{RGB}{255,230,230}
\definecolor{lightgreen+}{RGB}{210,255,210}

\begin{figure*}[t]
  \centering
  \resizebox{\textwidth}{!}{%
    \begin{tikzpicture}[
        node distance=0.9cm and 1.1cm,
        phase/.style={
          draw=black, rounded corners,
          minimum height=1cm, minimum width=1.7cm,
          inner xsep=2pt, inner ysep=2pt,
          align=center,
          font=\small\sffamily % we can use \footnotesize to make it a bit smaller
        },
        greyphase/.style={
          phase,
          fill=gray!20
        },
        darkgreyphase/.style={
          phase,
          fill=gray!40
        },
        arrow/.style={-Stealth, thick}
      ]

      % Nodes
      \node[phase] (rlci1) {RLCI\\inference};
      \node[greyphase, right=of rlci1] (rlc1) {RLC\\checking};
      \node[darkgreyphase, right=of rlc1] (res) {Code\\transformation};
      \node[phase, right=of res] (rlci2) {RLCI\\inference};
      \node[greyphase, right=of rlci2] (rlc2) {RLC\\checking};
      \node[greyphase, right=of rlc2] (rlfixer) {Enhanced\\RLFixer};
      \node[darkgreyphase, right=of rlfixer] (rlc3) {Patch\\validation};

      % \draw[arrow] ([xshift=-1cm]rlci1.west) -- (rlci1.west)
        % node[above, font=\scriptsize, near end, xshift=-7pt] {1};

      \draw[arrow] (rlci1) -- node[above, font=\scriptsize] {2} (rlc1);
      \draw[arrow] (rlc1) -- node[above, font=\scriptsize] {3} (res);
      \draw[arrow] (res) -- node[above, font=\scriptsize] {4} (rlci2);
      \draw[arrow] (rlci2) -- node[above, font=\scriptsize] {5} (rlc2);
      \draw[arrow] (rlc2) -- node[above, font=\scriptsize] {6} (rlfixer);
      \draw[arrow] (rlfixer) -- node[above, font=\scriptsize] {7} (rlc3);

      % \draw[arrow] (rlc3.east) -- ++(.8cm,0)
      %   node[above, font=\scriptsize, near end] {8};

      % \node[draw, align=left, font=\scriptsize, below=1cm of res, anchor=north] (legend1) {
      %   \textbf{Component Legend:} \\
      %   White box: reused component \\
      %   Grey box: enhanced component \\
      %   Dark grey box: new component
      % };
      \node[draw, rounded corners, thick, fit=(rlci1) (rlc3), inner sep=8pt, label=above:{\textbf{\OurTool}}] (arodnapbox) {};

      \draw[arrow] ([xshift=-1cm]arodnapbox.west) -- (arodnapbox.west)
        node[above, font=\scriptsize, near end, xshift=-7pt] {1};

      \draw[arrow] (arodnapbox.east) -- ++(.8cm,0)
        node[above, font=\scriptsize, near end] {8};

      \node[align=left, minimum width=3cm, font=\footnotesize, below left=1.3cm and .9cm of arodnapbox.west, anchor=west] (legend1) {
        \textbf{Inputs and Outputs:} \\
        1. Input Source Code, 2. Inferred Specifications, 3. RLC Warnings,
        4. Transformed Code, 5. Updated Specifications, 6. Updated Warnings, 
        7. Patched Code, 8. \revised{Validated Patched Code}
      };
      % Renaming 8. Statically & Dynamically Validated Patched Code makes the figure too small

      % \draw[arrow]
      %   (rlci2.north) -- ++(0,0.8)
      %   -| node[above, font=\scriptsize, pos=0.25] {inferred specifications} (rlfixer.north);

    \end{tikzpicture}
  }
% \todo{The labels for ``inputs and outputs'' are incorrect.  For example,
%   \#2 and \#3 include the source code, but they (and other labels) don't
%   mention the source code.  I think these slightly incorrect labels do more
%   harm than good, and therefore I suggest removing all the labels and
%   leaving the arrows unlabeled.}
  \caption{\OurTool's leak-repair pipeline.
% \todo{I don't think this text adds much.  It just makes the caption harder
  % to read:
% The flow begins with source code analysis using RLCI and RLC to extract
%   specifications and warnings.
% Code Transformation applies lightweight code transformations that improve repairability, after which \ourTool reanalyzes the updated code to generate refreshed specifications and warnings.
% RLFixer consumes both the warnings and the inferred specifications to produce code repairs.
% Final validation re-runs RLC on the patched program to ensure correctness.
Dark gray boxes indicate new components we introduced;
light gray boxes are existing components we extended;
white boxes are existing components we reused.
\mde{It might be nice to explicitly relate the components of this
architecture to the list of our contributions, showing where each is instantiated.}
% \todo{Move ``inputs and outputs'' text block left a bit, to avoid hanging
%   text at far right of figure.}
% \todo{Martin: The font is very small - small enough that I worry it will not be readable. Can we make it bigger? Do we really need labels on all of the arrows, or could we use some other kind
% of visual language to communicate the different kinds of communication?}
% \todo{Martin: I am not convinced that we need the arrow from RLCI to RLFixer. I would rather this diagram focus on control flow rather than I/O, personally.}
% \todo{Why is ``RLFixer'' not ``Enhanced RLFixer''? The text refers to it as such.}
% \todo{SM: I'm working on the figure.}
}
\label{fig:overview-pipeline}
\end{figure*}

% LocalWords:  xsep ysep greyphase darkgreyphase rlci1 RLCI rlc1 rlci2 sep
% LocalWords:  rlc2 rlfixer rlc3 arodnapbox xshift legend1

% By default, RLFixer only recognizes cleanup methods named \<close()> and does not apply nor verify its hints. While this ensures that no unsound patches are introduced, it also limits RLFixer's
% effectiveness in more complex or unconventional resource-handling
% patterns.
While RLFixer repairs some leaks, it struggles with
many real-world code patterns. It cannot handle any leak caused by reassignment{\,---\,}overwriting a field without first closing the resource{\,---\,}since it cannot guarantee safe cleanup when the
resource may escape the current scope. Its ability to reason about wrapper
classes is limited to cases where the class explicitly defines a method
named \<close()>.
%  and meets other heuristic match rules: a resource is passed to the class's constructor, stored in a field, and that field is closed in \<close()>\mde{What are they?
  % Explain or forward reference. SM: Need to discuss about mentioning this rules. I removed the heuristic rules as they are meant for wrapper resource aliases, not resource wrappers. RLFixer's rules focus on resource wrappers that take an incoming resource, while resource wrappers can also be declared without an incoming resource}.
In some cases, RLFixer's alias reasoning concludes a resource may escape when in fact it does not, thereby missing fix opportunities.
%Repairs are skipped when aliasing or delayed initialization prevents safe classification.
%Furthermore, 
% \todo{(Manu) somewhere maybe we want to bring out more that many of RLFixer's limitations relate to resources stored in fields, and many of our new techniques address that directly, either by shifting leak warnings to wrappers or doing more precise escape analysis.}

About 62\% of leaks that RLFixer cannot repair
in the NJR benchmarks~\cite{palsberg2018njr} involve resources stored in
fields.
\mde{Explicitly relate these to the list of contributions in the
  introduction.  The introduction contained an explicit list of 3 numbered improvements.}
\OurTool's techniques address
exactly this scenario:
ownership inference moves the leak warning from the field assignment back to
the wrapper allocation, and a new pre-close insertion repair, guided by a more
precise escape analysis, lets RLFixer safely close resources even when the field is reassigned.

% These limitations motivate the enhancements presented in the next section. Our goal is to extend RLFixer with inference-aware reasoning, apply lightweight source transformations to expose resource
% ownership more clearly, and add new fixes and repairs for field overwrites\,---\,ultimately enabling a more complete repair pipeline that works on unannotated, real-world code.

% LocalWords:  RLC RLFixer MyWriter PrintWriter pw unrepairable RLCI NJR

\section{Example Repair}
\label{sec:mot-example}

% \todo{(Manu) I feel this example does not show off the sophistication of our work enough and makes it look simplistic.  I suggest we say it is based on code seen in NJR examples but modify it.  We should rename \<GenerateDeterminantFromMinor> to something shorter.  We should add another method like \<resetStream()> to the class that overwrites the \<stream> field, so it cannot be made final.  And we should show how our technique can also repair the \<resetStream()> method by closing before overwriting.  Probably we should also show at least some of the inferred annotations in part (b) as well.  This will take more space so we need to be a bit careful.}

% \todo{\Cref{fig:mot-example} must appear on the same page as this text in the version of the paper that we submit. SM: I can't seem to do it. If I move the example to this page, then the SEC III, goes to the previous page.  Manu: I couldn't do it either.  Because it's a figure* the usual things to force figure placement don't seem to be working.  If we really want this we may need to split it into two single-column figures or something. Martin: I also tried to mess around with this, and couldn't find a solution that keeps the same general structure of both the figure and the text. It's not that important, so it's okay if we give up on this.}

% In case we want more detailed footnote.
% The full file path is url882f91ec97_WenboCao_Microsoft_Drone_tgz-pJ8-org_ejml_alg_dense_decomposition_qr_StabilityQRDecompositionJ8/src/org/ejml/alg/dense/misc/GenerateDeterminantFromMinor.java
\Cref{fig:mot-example} is a simplified but representative real-world
resource leak from the NJR dataset~\cite{palsberg2018njr}
that RLFixer cannot repair without
\ourTool.
Class \<TempFileWriter> allocates a \<PrintStream> both in its
constructor and in the \<resetStream> method.  RLC reports leaks at both
these locations.  The \<PrintStream> objects are written into fields and
thus might escape the current scope, so RLFixer deems both warnings
unrepairable and emits no fix.  Further, RLC inference (RLCI) cannot infer
that \<TempFileWriter> is a wrapper type, because the class does not
implement a finalizer method like \<close>.  

% On this code RLC reports two leaks: one in the constructor (instead of at the \<TempFileWriter> allocation in \<print>) and one in \<resetStream>.
% The class does not
% implement a finalizer method like \<close>, so inference cannot
% mark the stream as owned.  \todo{This sentence came out of the blue.  I
%   assumed, from ``RLFixer cannot repair without
% \ourTool'', that this paragraph is about RLFixer in the absence of
% \ourTool.  So why is inference mentioned at all in this paragraph?}

\emph{Injecting Finalizer to Enable Ownership Inference.}
\OurTool injects
a \<close()> finalizer and the \<AutoCloseable> interface on
\<TempFileWriter> (shown in blue in \cref{lst:mot-example-fixed}) to aid RLCI.
These changes
enable RLCI to add an \Owning annotation on the \<stream> field, explicitly marking ownership.
Because the class now owns the resource, RLC
now reports
a leak at the \<TempFileWriter>'s allocation inside
the \<print> method instead.

\label{def:shifted-leak-warning}
The new warning inside \<print> is a \emph{shifted leak warning}.
\revised{A shifted leak warning occurs when RLC reports the same defect at a different program location under a different analysis configuration (e.g., after adding ownership information). The underlying defect is unchanged; only the report location shifts.}
Here, the added \<close> method does not fix a leak, but
is needed for repair.  The \<PrintStream> created in the \<TempFileWriter>
constructor \emph{cannot} be closed there, as it may be used later by the
\<printSomething> method.  The resource cleanup \emph{must} instead occur
at the end of the enclosing \<TempFileWriter>'s lifecycle.\looseness=-1

\emph{Fix Generation.}
Given these transformations, RLFixer can generate repairs. As shown in green in \cref{lst:mot-example-fixed},
it encloses the allocation in \<print> within \tryfinally.   Further, our
new \emph{pre-close insertion} repair (\cref{sec:preclose}) inserts a pre-close
that closes \<stream> before reassignment (line 9).

This example reflects a common pattern in real-world Java code: user-defined wrapper classes that manage
resources internally.
Two types of mistakes can lead to resource leaks.
Resources may leak because the wrapper class has mistakes (like the missing \<close()>
method in this example, although mistakes in real code that \ourTool can
fix are often much subtler!).
Resources may also leak because the wrapper class is
misused (as in the \<print()> method in class \<Client>).
Our system makes both cases analyzable and repairable
with minimal structural edits, so that automated tools can handle patterns that required manual annotation and refactoring before.\looseness=-1

% LocalWords:  NJR unrepairable GenerateDeterminantFromMinor RLC RLFixer
% LocalWords:  url882f91ec97 WenboCao TempFileWriter resetStream RLCI
% LocalWords:  printSomething

\section{Detection and Repair Architecture}

The \ourTool repair pipeline (\cref{fig:overview-pipeline}) integrates static leak analysis (RLC), inference (RLCI), code transformation,
and repair (RLFixer).
It begins by combining RLCI and RLC to surface both inferred \MustCall
obligations on user-defined classes and resource leak warnings.
Leveraging these outputs, the pipeline applies a suite of lightweight \emph{Code Transformations}{\,---\,}marking eligible fields \<final>, converting
fields to local variables, and injecting missing \<close()> finalizers into wrapper classes (\cref{sec:implementation-code-transforms}).
It then re-invokes inference and RLC to produce updated, more actionable leak warnings.

An \emph{enhanced RLFixer} consumes the
improved warnings and specifications
from the transformed code to synthesize concrete code patches (\cref{sec:patch-materialization}).
(The original RLFixer requires a developer to \emph{manually} apply its fix hints.)
%\todo{There is no pipeline section corresponding to \cref{sec:implementation-field-overwrite}. Why not? The section should be organized in such a way that there
%  is something sensible to say here; if that means that \cref{sec:implementation-field-overwrite} needs to be merged into another section, so be it.}
% After running the enhanced RLFixer, \ourTool applies the synthesized fix
% hints as concrete code patches
% (\cref{sec:patch-materialization}),%
% %\todo{This does not appear in the pipeline.}
% meaning that \ourTool can be deployed in a fully-automated way, unlike RLFixer (which requires a developer to \emph{manually} apply its fix hints).
% %
% Finally, \ourTool re-runs RLC on the patched program, to ensure that the repairs addressed the corresponding warnings.

Using RLCI to infer ownership annotations leads to a new class of warnings (about possible overwrites
of fields containing resources) that the original RLFixer did not need to consider. We enhanced
both RLC itself and RLFixer to allow \ourTool to handle these warnings; as
this problem is logically distinct,
\cref{sec:implementation-field-overwrite} describes the necessary
modifications separately.

Finally, \ourTool validates the patches with both static and dynamic checks (\cref{sec:patch-validation-description}):
it recompiles, re-runs RLCI+RLC to ensure the warning is eliminated, and runs the project's JUnit tests to confirm test outcomes are preserved.

%% This section presents the architecture of \ourTool.  In \S\ref{sec:implementation-overview} we give a high-level overview of the leak detection and repair pipeline,
%% tracing how inferred specifications and warnings flow across multiple analysis and repair stages.
%% \S\ref{sec:implementation-code-transforms} details the lightweight \emph{Code Transformations} that reshape resource usage patterns for repair. 
%% In \S\ref{sec:implementation-rlfixer-enhancements} we outline the \emph{Enhancements to RLFixer}, such as support for inferred finalizers and \rca.
%% \S\ref{sec:implementation-field-overwrite} describes our handling of field-overwrite warnings, including false-positive filtering and safe pre-close insertion.
%% Finally, \S\ref{sec:implementation-llm-patch} covers our patch
%% generation and the RLC-based verification stage.

\subsection{Code Transformations}
\label{sec:implementation-code-transforms}
Static code transformations play a central role in our pipeline by
reshaping code to improve analyzability and fixability. These
transformations preserve program semantics while
clarifying ownership structure and exposing resource lifecycles. Each
transformation is designed to be sound and grounded in software engineering best
practices.

\subsubsection{Field Transformations for Immutability}
Mutable resource-holding fields introduce risks in resource management. If a resource field is reassigned without closing the previous value, the original resource may become
unreachable, causing a leak. By marking some such fields as \<final> and
making other fields local, \ourTool eliminates the possibility of
reassignment and makes lifecycle tracking simpler and safer.
\Cref{sec:implementation-field-overwrite} describes further
enhancements to RLC and RLFixer for cases where these transformations do not apply.\looseness=-1

\textbf{Preventing Reassignment}
\revised{\OurTool adds the \<final> modifier to a private field if the
  field is only assigned once.  For instance fields, the assignment must occur
  at object
  % This is intentially the Java 8 version of the JLS, not a more recent
  % version, to show how long-standing the concept is.
  construction time \cite[\S 8.3.1.2]{GoslingJSBB2014}.
  This requires that all assignments to the field occur
  at its declaration (via an initializer), or within constructors,
  or within one instance initializer block (exactly one of the possibilities), and only once along
  any path.  
  This is closely related to the notion of \emph{effectively final} \cite[\S
  4.12.4]{GoslingJSBB2014}, though that is defined only for local
  variables.
  Static fields are similar.}
% \OurTool makes resource fields \<final> whenever possible.
When the field assignment occurs inside \trycatch, \ourTool introduces a
temporary variable (\cref{lst:final-try-catch}) to satisfy Java's final-field
assignment rules.
Use of \<final> aligns with standard principles to prefer immutability
%from functional programming~\cite{okasaki1999purely,birrell2012immutability}
and improves lifecycle tracking in tools like RLC and RLFixer. 

\revised{\textit{Validation:} This transformation is semantics-preserving: 
the Java compiler rejects subsequent writes to \<final> fields.
%% Mike agrees that this is not worth bringing up.
% \footnote{\manu{Technically final fields \emph{can} be overwritten using reflection, though they are trying to change Java to get rid of this; see \url{https://openjdk.org/jeps/8349536}.  If there is any issue here it's an (unlikely) unsoundness in RLC, not \OurTool, so I don't think we need to mention it.}}
\OurTool applies this transformation even without an RLC warning.
}

\begin{figure}
\begin{lstlisting}[language=diff]
-private ServerSocket serverSocket;
+private final ServerSocket serverSocket;

 public MyClass(int port) {
+  ServerSocket tempSocket = null;
   try {
-    serverSocket = new ServerSocket(port);
+    tempSocket = new ServerSocket(port);
   } catch (IOException e) {
     e.printStackTrace();
+  } finally {
+    serverSocket = tempSocket;
   }
 }
\end{lstlisting}
\caption{Using a temporary variable for final assignment}
\label{lst:final-try-catch}
\end{figure}

\textbf{Reducing Scope}
\revised{\OurTool converts a \emph{private} resource field into a method-local variable when a simple syntactic check finds an
unconditional assignment to the field that precedes any reads in that method (including inside \trycatch), 
and there is no externally callable setter that assigns to the field.}
% the field isn't effectively shared\mde{What does ``effectively shared'' mean?}
% across methods (e.g., via a trivial\mde{What is a ``trivial'' setter?} setter).}
% When a resource-holding field is only used within one method, \ourTool converts it into a local variable.
This shortens the variable's lifetime and improves precision in static ownership reasoning.
This transformation follows the principle of minimal variable scope~\cite{Bloch2001}.

\revised{
  \textit{Validation:}
  This transformation is sound except when the field is reflectively accessed.
  Because reflective accesses do occur in practice,
  \ourTool validates these changes using the static and
dynamic protocol in \S\ref{sec:patch-validation-description}. \OurTool applies this
transformation even when there are no RLC warnings.
}

\subsubsection{Adding Finalizers to Wrapper Classes for Ownership Visibility}\label{sec:insert-close}
Wrapper classes that hold resources in fields may erroneously fail to
define a public cleanup method like \<close()>, e.g., \<TempFileWriter>
in \cref{lst:mot-example-leaky}.
The absence of a cleanup method prevents RLC inference from inferring the field is owning, as its rule for inferring ownership requires an attempt to dispose of the resource field~\cite{ShadabGTEKLLS2023}.
%To address this,
\OurTool injects
a \<close()> method into each such class \<C> and adds \<implements AutoCloseable> (\cref{lst:mot-example-fixed}).
RLFixer can later insert calls to the method to fix leak warnings on \<C> objects.

% \begin{figure}
% \begin{lstlisting}[language=diff, caption={Injecting a finalizer into a wrapper class\todo{Do we need this listing, or is it duplicative with the motivating example?}}, label=lst:finalizer-injection, aboveskip=0.5em]
% -class MyWrapper {
% +class MyWrapper implements AutoCloseable {
%    private final PrintWriter out;

%    MyWrapper(String path) {
%      this.out = new PrintWriter(path);
%    }

% +  @Override
% +  void close() {
% +    if (out != null) {
% +      out.close();
% +    }
% +  }
%  }
% \end{lstlisting}
% \end{figure}

This transformation is only applied when a resource is allocated
within a constructor for class \<C> and then assigned to an instance field
within the same constructor.  We found that in practice, such code strongly
suggests that \<C> owns the resource, despite \<C> lacking a cleanup
method.  \OurTool discovers such constructors by parsing leak warnings
from RLC\@.  With the transformation, RLC inference determines the
field is \Owning, shifting responsibility for closing the
resource to clients of \<C>.

\revised{
  \textit{Validation:}
  Since the \<close()> method is new, no
  clients of \<C> will have invoked it previously.
  Declaring \<implements AutoCloseable> merely exposes a standard cleanup interface (enabling optional \<try-with-resources>).
% \mde{Discuss adding \<extends AutoCloseable>?}
%  Adding a new \<close()> method\manu{This discussion is about adding new \<close()> \emph{calls}, right?  I don't think adding a \<close()> method (but not any calls to it) can alter behavior?  I think we should say that injecting finalizers cannot alter behavior, and we discuss injecting calls to those finalizers later, maybe with a forward reference.} alters
% client-visible behavior, so we only apply this when there
% is a relevant RLC warning. \OurTool uses the static and dynamic protocol in \S\ref{sec:patch-validation}
% to validate the injected finalizer.
}

\subsection{Enhancements to RLFixer}
\label{sec:implementation-rlfixer-enhancements}
We extended RLFixer in two ways to improve its ability to repair leaks involving resource-holding fields (\cref{sec:inferred-finalizers,sec:containment}).
We also added support for concrete patch materialization (\cref{sec:patch-materialization}).

\subsubsection{Support for Inferred Finalizers}
\label{sec:inferred-finalizers}
RLC inference sometimes determines that an existing class method, which may have any name, is a finalizer and annotates the code accordingly.
However, RLFixer only supports adding calls to methods named \<close>.
To address these cases, we extended RLFixer with the ability to insert
calls to any finalizer method. The enhanced RLFixer parses inferred \MustCall annotations on user-defined classes to determine which
methods should be treated as finalizers. If a class \<C> is annotated with \MustCall\<("shut\-down")>, our extended RLFixer treats \<shutdown> as the finalizer method
for \<C> objects. This change broadens RLFixer's resource model: any class
with an inferred finalizer is now treated as a resource wrapper.
This allows RLFixer to produce repairs for
classes it would previously ignore.

\subsubsection{\RCA}
\label{sec:containment}
RLFixer avoids introducing use-after-close errors by checking if resources
\emph{escape} into fields or data structures before generating a
fix~\cite{UttureP2023}.
% \mde{``before applying a fix'', and the next
%   sentence, seem to contradict
%   earlier assertions that RLFixer gives up if a resource escapes.}
However, RLFixer allows a resource to be written into a field of an object
it determines to be a \emph{resource alias}.  An object is a possible
resource alias if its class contains a resource field, assigns to that
field in its constructor, and defines a cleanup method.  Identified
resource alias objects are then considered during RLFixer's escape analysis
to ensure that they themselves do not escape into other fields. 

\begin{figure}
\begin{lstlisting}[language=Java]
public class FileEventProxy {
  private Scanner scanner;
  public FileEventProxy(InputStream in) {
    this.scanner = new Scanner(in);
  }
  public boolean hasNextEvent() {
    return scanner.hasNextLine();
  }
  // other uses
}
\end{lstlisting}
\caption{A resource accessor class uses, but does not own or close, a resource.}
\label{lst:containment-original}
\end{figure}

We identified two issues with RLFixer's escape reasoning.  First,
RLFixer is \emph{too conservative} in the presence of \emph{resource
accessor} objects that use a resource but do not take ownership of it.  A
resource accessor object takes a resource as a constructor argument and
uses it during its lifetime, but does not close the resource.  For example,
in \cref{lst:containment-original},
a \<FileEventProxy> object uses the \<InputStream> passed to its
constructor but does not close it.
RLFixer does not treat resource accessors as aliases of the underlying resource, since they cannot be used to close
the resource.
% Such resource accessors are not 
% \mde{Can we use active voice and a more specific verb than ``considered''?}
% resource aliases by RLFixer, since they cannot be used to close the resource.
So, RLFixer treats passing a resource into a resource
accessor as a field escape, preventing repair.  However, as long as a
resource accessor does not outlive the underlying resource, it should not
preclude a repair.

Second, RLFixer's checking is \emph{unsound}, as it does not check whether
objects may escape further via reads of resource alias fields.  For
example, consider a class \<Wrapper> with a field \<InputStream s> and a
method \<getStream()> that returns \<s>.  Even if \<Wrapper> meets
RLFixer's resource alias conditions outlined above, it may not be safe to
close the resource passed to \<Wrapper> in the original scope, since the
resource may leak further via \<getStream()>; RLFixer does not check
this condition.
We did not observe this unsoundness in the NJR
dataset~\cite{palsberg2018njr}, but addressing this issue remains important
for overall safety.
% \todo{Mention somewhere that we didn't observe unsoundness in practice on NJR but still important to fix}

% ~In the default RLFixer implementation, a class is only treated as a resource wrapper if it contains a resource field, assigns to that field in its constructor, and defines a cleanup method
% such as \<close()>. This logic fails to recognize a large class of wrappers that receive a resource handle externally (e.g., via constructor) and use it internally, but do not define
% an explicit finalizer. These classes encapsulate the resource but do not formally own it{\,---\,}i.e., they do not define a finalizer or carry ownership annotations.
% RLFixer won't repair them until we prove the underlying handle never outlives the wrapper.

To handle these issues, we introduced sound support for resource accessors
via \emph{\rca}, a lightweight extension to RLFixer's escape analysis.
\Rca checks that resources stored in a field do not further escape to some
long-lived data structure.  \Rca is used both to identify resource
accessors and to soundly check for resource aliases.  \Rca is again
used when introducing fixes for non-final field overwrites; see
\Cref{sec:preclose}.

% wrapper-identification logic that lets us recognize classes as resource owners\todo{they are not owners, they are borrowers}
% even without a finalizer.  By combining alias tracking with a targeted escape check, we prove that the wrapper's internal field never propagates into any long-lived data structure,
% allowing RLFixer to treat the class as a valid wrapper and generate client-side cleanup hints.

% To specify our \rca, we first define \rc:
\begin{definition}[\Rc]
\label{def:containment}
An instance field $f$ of a class $C$ is \emph{contained} iff for every $C$ object $c$, there is no data flow from $c.f$ into any field, array, or data structure whose lifetime may exceed that of $c$.
% Let \(C\) be a Java class and \(f\) a \<private> field of \(C\) assigned exactly once in \(C\)'s constructor from an external handle.  We say \(f\) is \emph{contained} if,
% after that initial assignment, an interprocedural escape analysis (over all uses of \(f\)) finds \emph{no} paths by which the handle in \(f\) flows into any field, array, or
% data structure whose lifetime may exceed that of \(C\). 
\end{definition}
Our analysis conservatively checks for \rc.  We require the field $f$ is private, to limit the initial scope of analysis to $C$.  Then, for each read of $f$ within $C$, we check if the value read from $f$ may flow into some field, array, or data structure, re-using the extant RLFixer escape analysis~\cite{UttureP2023}.  

Given our \rca, we enhanced RLFixer's identification of resource aliases as
follows.  During resource alias identification, the enhanced RLFixer also runs \rca to ensure
that further escapes from the field of the resource alias are not possible.
Further, if a class meets all requirements for being a resource alias but lacks a finalizer method (which can be \<close> or any method inferred via \MustCall),
% \todo{only \<close>? Or can this be any finalizer?}
our enhanced RLFixer categorizes it as a resource accessor.  We enhanced RLFixer's escape
analysis to treat resource accessor objects identically to
resource aliases, except that a resource accessor cannot be used to close a
resource.

% To make use of containment, RLFixer invokes its interprocedural escape-analysis (now over the set of \emph{resource aliases}) on each candidate field \<f>'s uses. If Definition
% \ref{def:containment} holds, we mark \<C> as a valid resource wrapper{\,---\,}even without finalizer method (e.g., \<close()>) and proceed to repair its client sites.

% \paragraph{\RCA Algorithm.}  
% For each private resource field \(f\) in class \(C\):

% \begin{enumerate}[nosep]
%   \item \textbf{Alias Collection.}  Identify every use of \(f\) and compute all its aliases.
%   \item \textbf{Escape Scan.}  For each use of \(f\) or its aliases, check whether the handle is ever written into any other field or stored into an array or collection{\,---\,}such
%   a write indicates the resource can outlive \(C\).
%   \item \textbf{Containment Verdict.}  If no escaping write is found, declare \(f\) contained; otherwise, containment fails and \(C\) is not treated as a wrapper.
% \end{enumerate}
% This check leverages RLFixer's existing alias and dataflow infrastructure.  

% \paragraph{Concrete Instance.}  
% By default, RLFixer ignores classes like \<FileEventProxy> above because they lack an explicit \<close()> method and therefore do not qualify as resource wrappers. We also avoid naively injecting a
% synthetic finalizer{\,---\,}without clear evidence that \<FileEventProxy> intends to manage the resource's lifecycle, adding a \<close()> risks over-instrumentation.

As an example, suppose a leaking resource is stored in a \<FileEventProxy> object (\cref{lst:containment-original}).  With RLFixer's original logic, this would prevent a leak repair due to a field escape.  With our enhancement, RLFixer can soundly show that the object is a resource accessor, as \rca on the field \<scanner> proves the resource does not leak further via the field.  Then, as long as the \<FileEventProxy> object does not escape, the repair can be applied, as seen in \cref{lst:containment-diff}.
\begin{figure}
\begin{lstlisting}[language=diff]
  InputStream s = new FileInputStream("file.txt");
  FileEventProxy proxy = new FileEventProxy(s);
+ try {
    // use proxy
+ } finally {
+   s.close();
+ }
\end{lstlisting}
\caption{Fix for a client of \cref{lst:containment-original}, enabled by handling of resource accessors.}
\label{lst:containment-diff}
\end{figure}

An alternative to introducing resource accessors would have been to insert a \<close> method into \<FileEventProxy> (\cref{sec:insert-close}), making it a resource alias.  However, such a change is unnecessary to repair the leak: the resource is passed into the constructor, so the client can close it (contrast with \cref{lst:mot-example-fixed} where the resource is allocated in the constructor).  Adding support for resource accessors leads to less intrusive changes, avoiding insertion of unnecessary \<close> methods into types that do not require them.
% Now, RLFixer generates a client-side cleanup hint at each call site{\,---\,}i.e.\ where \<new FileEventProxy(stream)> is invoked{\,---\,}by inserting a \<try/finally> that closes the wrapped \<InputStream>.
% This approach closes the incoming handle exactly once, soundly repairing the leak without altering \<FileEventProxy> itself.
% Because containment guarantees that \<stream> never leaks beyond \<FileEventProxy>, closing it in the client's \<finally> block is both sound and sufficient.  

% \todo{(Manu) I think we can cut this next subsection for space if needed}

% \subsubsection{Improved IR Matching}
% ~RLFixer relies on mapping RLC warnings\,---\,reported at the Java source level\,---\,to internal WALA IR instructions for performing escape analysis and fix generation. However, its original matching
% logic was fragile in the presence of nested or anonymous classes, often failing to resolve these warnings to valid IR locations. As a result, a nontrivial portion of reported leaks were excluded
% from further analysis.

% We improve this mapping logic to support a broader range of structural cases, allowing RLFixer to process warnings originating from complex class hierarchies and non-top-level declarations.
% This change does not alter the repair logic itself, but significantly increases the number of actionable warnings that reach the analysis phase.

\revised{
\subsubsection{Patch Materialization}
\label{sec:patch-materialization}
While RLFixer produces structured, parameterized repair hints, they are
purely textual, e.g.,
% \mde{Make this a concrete example, without templating  or metavariables.}

\begin{quote}
  Add following code below line \<22> (WriterFile.java): \codeid{finally\{ try\{
    <NEW\_VARIABLE>.close(); \} \}} \\// where variable \codeid{<NEW\_VARIABLE>} points to
  the resource from line \<17>.
\end{quote}
  
\noindent
\OurTool extends RLFixer with a deterministic \emph{patch materializer} that takes (i) the RLC warning, (ii) the corresponding RLFixer hint, and (iii) the relevant source,
and then (a) parses the compilation unit into an AST, (b) locates the edit sites indicated by the hint, and (c) applies template-guided AST rewrites to realize the repair
(e.g., wrapping with \tryfinally or try-with-resources when legal, or inserting a \<close()> call).
The modified AST is pretty-printed and diffed against the original file to produce a concrete patch.
}

\subsection{Field Overwrite Handling}
\label{sec:implementation-field-overwrite}

RLC issues an \textit{owning field overwrite} warning
when a resource field is reassigned without first being closed.
The assignment potentially causes a leak by making the original resource
unreachable. \OurTool addresses this issue in two ways: by eliminating
false positives
in safe constructor-based assignments, and by safely inserting closure logic before actual reassignments.
The original RLFixer{\,---\,}which runs RLC without any annotations{\,---\,}did
not need to consider this case.
RLC only issues these warnings if at least one field is annotated as
\Owning (in \ourTool, by inference).

\subsubsection{Filtering False Positives on Constructor Assignments}\label{sec:constructor-fp}
Previously, RLC would report an overwrite warning for any write to a non-final resource field in a constructor, even when it was the first write to the field. We improved RLC
to not issue a warning if 6 conditions apply.
The field (1) is private,
(2) has no initializer at its declaration, and (3) is not written in any
instance initializer block.
(4) The assignment occurs directly in the constructor body.
The constructor (5) writes the field exactly once and
(6) neither delegates via \<this(...)> nor performs any
method calls before the assignment.
These constraints conservatively guarantee, based on Java's initialization order, that the assignment is the first write to the field, not an overwrite.  The first write to a field cannot cause a leak since the field has no previous value.

% Since a resource cannot be leaked
% before it is assigned, reporting a leak in this case is a false positive. Filtering such warnings reduces noise in the leak warnings and improves the signal quality passed on to repair stages.

\subsubsection{Safe Reassignment Fixing via Pre-Close Insertion}
\label{sec:preclose}
For cases where a resource field \(f\) is legitimately reassigned outside
a constructor, we devised a new repair that inserts a conditional close of
the field's current value just before the reassignment, thereby preventing
a leak.  Care must be taken to ensure that this repair does not introduce a
use-after-close error due to some other outstanding pointer alias for the resource.  \OurTool only applies this transformation
when the following conditions hold:
\begin{enumerate}
  \item \(f\) is a \<private> field of the enclosing class \(C\).
  \item All writes to \(f\) assign it a newly-allocated resource (e.g., \<new Socket(...)>).
  \item \Rc (Def.~\ref{def:containment}) holds for \(f\), i.e., the resource never escapes the class from \(f\).
\end{enumerate}
Condition 1 limits the scope of analysis, while conditions 2 and 3 ensure
there cannot be other references to the resource that outlive the method re-assigning the field.  When these conditions are
met, enhanced RLFixer safely inserts the repair shown in \cref{lst:preclose}.
\begin{figure}
\begin{lstlisting}[language=diff]
+  if (socket != null) {
+    try {
+      socket.close();
+    } catch (IOException e) {
+      e.printStackTrace();
+    }
+  }
   socket = new Socket();
\end{lstlisting}
\caption{Example repair for field overwrite.}
\label{lst:preclose}
\end{figure}
This fix closes the previously held resource before overwriting it, preventing a leak.  It is possible that the resource was already closed, and hence the inserted call is a duplicate.  But, this does not cause problems in practice, because Java close methods are typically specified to be idempotent and hence safe to repeat (see, e.g., the \<java.io.Closeable\#close> documentation~\cite{closeableCloseDocs}).
% \todo{Discuss what happens if \<close> can throw an exception}
This repair template simply prints the stack trace of any exception thrown by \<close()>, but the behavior in the catch block could easily be customized (e.g., to re-throw the exception or perform logging).

% ensuring the reassignment always occurs; if \<close()> fails, the leak persists{\,---\,}a rare but acceptable risk.

\revised{
\subsection{Patch Validation: Static and Dynamic}
\label{sec:patch-validation-description}
\OurTool validates each materialized patch with two gates:

\subsubsection{Static Validation.}
The project must recompile cleanly, and a re-run of RLCI+RLC on the patched code must confirm that the originally reported leak at the patched site is eliminated.

\subsubsection{Dynamic Validation.}
\OurTool executes the project's JUnit test suite and checks that no previously passing test fails post-patch (i.e., no new regressions relative to the pre-patch baseline).
}

\section{Implementation}
\OurTool builds on the Checker Framework's Resource Leak Checker (RLC)~\cite{KelloggSSE2021}.
The Checker Framework distribution includes RLCI~\cite{ShadabGTEKLLS2023}.
We wrote Error Prone~\cite{ErrorProne} plugins for field transformations: converting fields to be final and converting resource fields to locals.
\OurTool injects wrapper-finalizers using JavaParser~\cite{JavaParser}.
For static analysis, \rc and escape analyses are run on WALA's SSA intermediate representation~\cite{WALA}.
We also fixed RLFixer's source-to-IR matching logic to correctly map warnings within nested and anonymous classes, enabling more repairs.
% \revised{
% Patch synthesis is performed by extracting only the relevant source file containing the leak and its RLFixer hint,
% which are then processed by a deterministic AST transformer implemented with JavaParser.
% }
The full pipeline targets Java 11.
% Retain this, even though the URL also appears elsewhere.  Don't make
% readers hunt for it or wonder why it isn't mentioned here.
% The complete artifact, including source code and benchmarks, is available at
% \href{https://doi.org/10.5281/zenodo.15542577}{10.5281/zenodo.15542577}.
%% \todo{This URL doesn't work for me. DOIs and Zenodo are usually not anonymous, either, so I'm not sure why we are
%%   using them for a submission - they are appropriate for a camera-ready, but in a submission it is critical that we
%%   not reveal who we are to the reveiwers! SM: It should work now. My docker image became too large, so I couldn't use github like websites. I've maade the publisher anonymouse in Zenodo, so it should be fine.}

% LocalWords:  RLCI WALA's LTS vCPUs Zenodo DOI

\section{Experimental Setup}
Our evaluation of \ourTool in detecting and repairing resource leaks
aimed to answer the following research questions:

\begin{researchquestions}
    \item How effective is \ourTool in reducing and repairing leak warnings
      compared to RLC+RLFixer?
    \item How effective is \ourTool compared to RLCI+RLC+RLFixer (adding inference~\cite{ShadabGTEKLLS2023} to the RLC+RLFixer combination, without our other improvements)?
    \item How much do the different components of \ourTool contribute to its effectiveness?
    \item For leaks that cannot be repaired by \ourTool, what is the root cause?
\end{researchquestions}

\subsection{Dataset}
The evaluation uses 285 of the 293 Java 8 projects in the NJR-1 dataset~\cite{palsberg2018njr}, with each project averaging 6,028 non-blank, non-comment lines of Java code.
8 projects are excluded due to timeouts during RLC inference.
This same benchmark was used to evaluate RLFixer~\cite{UttureP2023}, though they did not exclude any projects
because they did not run inference.  The projects cover a wide range of domains.
% The main \ourTool pipeline (\cref{fig:overview-pipeline}) always took less than one hour per benchmark,
% excluding patch generation and final validation, making it suitable for batch (i.e., non-interactive) deployment;
% since the output of \ourTool is a effectively a pull request, batch deployment is reasonable.
% We omit patch generation time because it involves asynchronous LLM API calls that are not consistently measured.

\revised{The NJR-1 benchmark is an unlabeled snapshot of open-source Java
  projects; it does not contain ground-truth defects annotations.
  Our evaluation depends on resource-leak warnings reported by RLC, which
  is a sound tool.
  Many files have no resource leaks; our evaluation shows that \ourTool not
  only repairs leaks but does not break files that have no leaks.}

\subsection{Configurations}
\label{sec:eval-configurations}
We evaluated three configurations to answer RQ1 and RQ2:

\begin{enumerate}[nosep]
    \item \textbf{\RLCRLFixer}: RLC without resource specification
      inference and the original RLFixer without our enhancements (for RQ1).
    \item \textbf{\RLCIRLCRLFixer}: RLC with specification inference but unmodified RLFixer (for RQ2).
    \item \textbf{\OurTool}: As in \cref{fig:overview-pipeline}.
\end{enumerate}

\noindent All configurations are evaluated after patch materialization (\cref{sec:patch-materialization}) and
validation (\cref{sec:patch-validation-description}), for fairness.
% All leak and fix counts reported in \cref{tab:main-results} and \cref{tab:ablation-results} are
% computed after patch materialization (\cref{sec:patch-materialization}) and validation (\cref{sec:patch-validation-description})
% to ensure fair comparison.
% \cref{sec:patch-validation-experiment} shows that over 99\% of \OurTool patches pass validation.
% \manu{I massaged the previous sentence for the major revision deadline; please check it.}
% \mde{That seems wrong to me:  we should measure after repair validation. SM: Patching and validation is time-consuming (due to manual patching in some cases and patch conflicts too while merging. Doing it across six configurations will take a lot of time. So, we decided to do it only for the main configuration.)}

% \subsection{Fixability}

% For each configuration, we record all leak warnings reported by RLC.
% Each warning is processed by RLFixer, which first maps the warning to its corresponding SSA instruction in the program representation and performs escape analysis to check safety.
% If a repair is deemed safe, RLFixer generates a fix hint. Warnings that pass both mapping and safety checks are counted as \emph{fixed}.

\subsection{Weighted Fix Count}\label{sec:weighted-fix-count}

RLC inference and \ourTool's code transformations often cause leak warnings
to shift from uses of library resources to uses of corresponding wrapper
types, as discussed in \cref{sec:inference} (see the discussion of
\cref{lst:inference-example}).  While this shifting makes the leak warnings
more actionable and useful, from an experimental point of view it makes
comparisons across our configurations difficult, since a single warning on
a library resource in one configuration could be shifted to multiple
warnings about wrapper type objects in another.  Consistent with Shadab et
al.~\cite{ShadabGTEKLLS2023}, we see an \emph{increase} in the number of
total leaks when inference is enabled (see \cref{tab:main-results}).

% For example, what was once a single warning in a constructor may, after injecting annotations and finalizers, become several
% warnings at the various sites where the wrapper is instantiated. See \cref{fig:inference-example} for an illustration.

\begin{figure}[t]
  \centering\small
  \resizebox{\columnwidth}{!}{%
  \begin{tabular}{c|c|c|c|c|c|c}
    \toprule
    \multirow{2}{*}{\lower0.5\normalbaselineskip\hbox{\textbf{Configuration}}}
      & \multicolumn{3}{c|}{\textbf{Leak warnings}}
      & \multicolumn{2}{c|}{\textbf{Fixed warnings}}
      & \multirow{2}{*}{\mbox{\begin{varwidth}{1cm}\textbf{Repair rate}\end{varwidth}}} \\ 
    \cmidrule(lr){2-4}\cmidrule(lr){5-6}
      & \textbf{CL} & \textbf{XE} & \(\mathbf{XR}\)
      & \(\mathbf{F_{CL}}\) & \(\mathbf{F_{XE}}\)
      &  \\ 
    \midrule

    % ── RLC+RLFixer ──
    \multirow{2}{*}{\lower0.5\normalbaselineskip\hbox{\RLCRLFixer}}
      & \multicolumn{3}{c|}{\BOneTot}
      & \multicolumn{2}{c|}{\BOneFixTot}
      & \multirow{2}{*}{\lower0.5\normalbaselineskip\hbox{\BOneR\%}} \\ 
    \cmidrule(lr){2-4}\cmidrule(lr){5-6}
      & \BOneCL & \BOneXE & \BOneXR
      & \BOneFixCL & \BOneFixXE
      & \\ 
    \midrule

    % ── RLCI+RLC+RLFixer ──
    \multirow{2}{*}{\begin{varwidth}{2cm}\RLCIRLCRLFixer\end{varwidth}}
      & \multicolumn{3}{c|}{\BTwoTot}
      & \multicolumn{2}{c|}{\BTwoFixTot}
      & \multirow{2}{*}{\lower0.5\normalbaselineskip\hbox{\BTwoR\%}} \\ 
    \cmidrule(lr){2-4}\cmidrule(lr){5-6}
      & \BTwoCL & \BTwoXE & \BTwoXR
      & \BTwoFixCL & \BTwoFixXE
      & \\ 
    \midrule

    % ── \OurTool ──
    \multirow{2}{*}{\lower0.5\normalbaselineskip\hbox{\OurTool}}
      & \multicolumn{3}{c|}{\BestTot}
      & \multicolumn{2}{c|}{\BestFixTot}
      & \multirow{2}{*}{\BestR\%} \\ 
    \cmidrule(lr){2-4}\cmidrule(lr){5-6}
      & \BestCL & \BestXE & \BestXR
      & \BestFixCL & \BestFixXE
      & \\ 
    \midrule
  \end{tabular}
  }
  \caption{Leak resolution breakdown across configurations. ``Fixed
    warnings'' values are \emph{weighted fix counts} (\cref{sec:weighted-fix-count}).}
  \label{tab:main-results}
\end{figure}

% LocalWords:  Config lr RLCI

To conservatively account for these differences, we define a \emph{weighted fix count}.
% \mde{\emph{Leaks} are never shifted.  Only \emph{leak warnings} are.
%   Furthermore, this definition comes much too late.  Please don't just find
%   a location in the text to add words to address a review comment.  Rather,
%   determine where in the text the confusion first arose and fix it there.
%   I did this on page \ref{def:shifted-leak-warning}.}
% \revised{A \emph{shifted leak} arises when inference and code transformations move a warning
% from one program location (e.g., a resource allocation in a wrapper constructor)
% to another location (e.g., a client instantiation of that wrapper). The underlying
% defect is unchanged; only the report location shifts.}
% For leaks on resource wrappers, the warnings are mapped back to their underlying resource origin\,---\,typically, the
% allocation of a library resource.
For each leak warning (see \cref{def:shifted-leak-warning}) on a library resource, we determine how many of its shifted warnings were successfully repaired. If $k$ out of
$n$ shifted sites are fixed, we assign a weighted fix score of $k/n$ to
that leak. \emph{Non-shifted} leak warnings (i.e., leaks reported on library resources) are
assigned a score of 1 if fixed or 0 if not.
The \emph{weighted fix count} ensures that each root library leak contributes in proportion to the fraction
of its associated warnings that are fixed, regardless of shifting due to inference and transformations.

In the general case, mapping a shifted leak warning back to a library leak warning can be quite challenging and require inter-procedural data flow analysis.  And for leak warnings on non-final field overwrites (\cref{sec:implementation-field-overwrite}), there may be multiple library resources possibly leaked by the overwrite.  In our experiments, we used a combination of automatic and manual analysis to compute the shifted leak warning mapping, and we separately categorized non-final field overwrites to avoid the complications of mapping those warnings.

% LocalWords:  NJR WALA enclosement Fixability unrepairable RQ1 RQ2 nosep
% LocalWords:  RLCI unenhanced

\section{Evaluation}

\subsection{Results}
\label{sec:eval-results}

% \subsubsection{Breakdown of leak warnings}
\newcommand{\Worig}{\ensuremath{\mathrm{W}_{\mathrm{orig}}}\xspace}
\newcommand{\Wxform}{\ensuremath{\mathrm{W}_{\mathrm{xform}}}\xspace}
% To make the effect of inference, code transformations, and repair passes
% explicit,\mde{This introduction makes it sound like the question should be
%   addressed by an ablation study, not by these statistics.}
% we partition leak warnings into three categories.
To compare different configurations (\cref{sec:eval-configurations}),
we partition leak warnings into three categories.
% THE NEXT LINE CAN BE ADDED IF NEEDED.
% by comparing the initial warnings with those generated after applying inference and code transformations.
%
Let \Worig be the warnings produced by RLC alone on
the original code.
Let \Wxform be the warnings produced by RLC
immediately before the enhanced RLFixer is run (at point 6 in
\cref{fig:overview-pipeline}), after mapping any shifted leak warnings (\cref{def:shifted-leak-warning})
back to their corresponding library leak warning.
Line-number changes do not affect whether two warnings are considered the same.

\begin{itemize}
  \item \textbf{Core Leaks (CL)} are $\Worig \cap \Wxform$. 
        % (Set operations are modulo changes in line numbers.) 
        These are warnings produced by RLC directly indicating a library resource or wrapper object is leaking.
        % \mde{How is this number computed?  Is it $\Worig \setminus
        %   \mathrm{XR}$?  Is it also equal to $\Wxform \setminus
        %   \mathrm{XE}$?  If those two values differ, what is our reason for
        %   choosing one over the other?}

  \item \textbf{Transformation-Exposed (XE) warnings} are $\Wxform
    \setminus \Worig$. These new warnings appear as a result
    of inference and transformation.  The dominant warning type in this
    category is overwrites of non-final \Owning fields
    (\cref{sec:implementation-field-overwrite}), where \Owning was added by
    RLCI inference.

  \item \textbf{Transformation-Resolved (XR) warnings} are $\Worig
    \setminus \Wxform$.  These warnings do not need repair: they
    were false positives that were fixed by semantics-preserving
    transformations or by adding resource specifications.
    To be treated as resolved, a warning must no longer appear \emph{and}
    no warnings on wrappers can be mapped to it
    (\cref{sec:weighted-fix-count}).  It is possible that the resource from
    a resolved warning could reappear as
    % \mde{Should ``leaks via'' be ``was shifted to''?  The current wording smacks of unsoundness.}
    a non-final field overwrite warning, but we
    manually inspected a sample of 50 \textbf{XR} warnings and never observed
    this to occur.

\end{itemize}

\noindent
The 
% \mde{I think this does not include warnings that occur at point \#3 in
  % \cref{fig:overview-pipeline} but not in \Worig or \Wxform.
  % Why are they not in the warning universe?}
warning universe is:
\[
  \mathrm{T} = \mathrm{CL} + \mathrm{XE} + \mathrm{XR}
\]

\noindent
and the \emph{resolution rate} is

\[
  \mathrm{R} = \frac{\mathrm{F}_{\mathrm{CL}} + \mathrm{F}_{\mathrm{XE}} + \mathrm{XR}}{\mathrm{T}}
\]

\noindent
where F\textsubscript{CL} and F\textsubscript{XE} denote the leaks actually fixed in each category.
% \todo{\Cref{tab:main-results} does not use the same italicization (math
%   mode) as used here.}

\Cref{tab:main-results} presents our main results.  For \RLCRLFixer, Utture
et al.~\cite{UttureP2023} reported a 51\% average fixable rate for RLC
warnings.  However, this calculation excluded RLC warnings that RLFixer could not
map to a WALA IR instruction, preventing repair.  Counting all reported
leaks, we found the actual fix rate for RLFixer for RLC warnings was
\BOneR\%.

% \mde{I moved the categorization to the end of this section, so discussions
%   of it need to be moved later as well.}

For \textbf{\RLCIRLCRLFixer}, which just adds RLC inference to \RLCRLFixer, \BTwoXE~new warnings are reported due to inference (the \textbf{XE} category), nearly all due to overwrites of non-final fields marked as \Owning by inference.  At the same time, inference leads to 357 of the original warnings being resolved, raising the
resolution rate from \BOneR\% to \BTwoR\%.  The resolution was significantly due to inference discovering 352 wrapper types, with 498 \Owning fields total.

%\todo{This paragraph surprisingly makes no mention of wrappers that inference detects, enabling new repairs. Given the story we're trying to tell, we should call out how many times that happens explicitly, rather than just attributing all of the improvement from inference to ``pruning''.}

\mde{I think that, throughout the paper, we should claim 1016 or fewer fixed warnings, rather than
  1025.  More generally, we should report fixed warnings after the ``Patch
  validation'' step of \cref{fig:overview-pipeline}.}

In \textbf{\OurTool} we observe a significantly larger \BestFixTot~leaks repaired (\BestFixCL~core and \BestFixXE~inference exposed) and \BestXR~warnings resolved, pushing the
resolution rate to \BestR\%.  \OurTool discovers 443 wrapper types and 627 \Owning fields.  This is a significant increase over \RLCIRLCRLFixer, due to our injection of \<close> methods.  At the same time, the raw \textbf{XE} count \emph{drops} from \BTwoXE~in \RLCIRLCRLFixer to \BestXE, despite the increase in the number of \Owning fields.  This decrease is due to the false-positive filtering of \cref{sec:constructor-fp}, which significantly reduces the number of \textbf{XE} warnings.
Overall, by exposing wrapper types and significantly enhancing repair capabilities related to wrappers, \ourTool resolves over two-thirds of all
reported leak warnings, significantly improving on other configurations.

Note that the weighted fix count metric (\cref{sec:weighted-fix-count}) used for \cref{tab:main-results} is intentionally conservative: that is, it \emph{understates} \ourTool's effectiveness
at repairing leaks of wrappers. In particular, consider a case where a warning about single library resource in a wrapper class is mapped to ten uses of that wrapper class.
In this case, if \ourTool fixed 5 of the 10 warnings, it would only get credit for fixing $0.5 = 5/10$ of a warning.
To illustrate \ourTool's effectiveness for wrappers more directly, in \RLCIRLCRLFixer, RLFixer could repair only 28 of 543 core leaks on wrappers (5\%); \ourTool repairs 412 of 838 warnings (49\%), an order-of-magnitude improvement in repair effectiveness.
% Notably, fixable \emph{shifted-unmapped} raw leaks climb from $28$ of $543$ in \RLCIRLCRLFixer to $412$ of $838$ in \ourTool{\,---\,}an order-of-magnitude
% gain that the headline weighted-fix metric only partially captures.

% \paragraph{Leak Count Discrepancies Across Configurations.}
% The total number of reported leaks increases from \RLCRLFixer to \RLCIRLCRLFixer due to new warnings introduced
% by inference. While we mapped user-defined wrappers and owning-field aliases across configurations, several
% categories of leaks are only reported with inference enabled. For example, RLC issues new warnings on non-final
% owning field overwrites even in cases where the original resource is not directly allocated by the class
% (e.g., a wrapper constructed with a static resource or a stream passed from another context). In some cases
%  RLC skips reporting the original allocation altogether and instead reports the leak on a field assignment
%  or call site, resulting in an unmatched leak between configurations. Other discrepancies arise when inference
%  triggers redundant warnings due to conservative must-call tracking\,---\,such as flagging a field as a leak despite
%  it being closed in most paths, or reporting multiple overridden field warnings when only one was emitted in 
%  the baseline. These issues collectively explain the increase in reported leaks under inference and why the total
%  leak counts do not exactly align across configurations.

\mde{Readers might want to see an end-to-end number.  That is, what is the
  number of RLC warnings on the original code, and what is the number of
  RLC warnings on the final code (possibly modulo shifted warnings)?  I
  think it may raise suspicion that the paper does not explicitly present
  either number. SM: To get the best final number, we need to run \ourTool in a loop. As RLC itself do not present all possible leaks in one run.
  I've noticed this while working on the implementation. After a patch has been applied, RLC reports new leaks that were not reported before.}

\subsection{Ablation Study}

% \begin{figure}[h]
% \centering
% \caption{Impact of disabling components in ablation study.}
% \label{tab:ablation}
% \begin{tabular}{lcc}
% \toprule
% \textbf{Configuration Variant} & \textbf{Weighted Fixed} & \textbf{Fix Rate} \\
% \midrule
% Full Pipeline (\OurTool)       & 1442                    & 68\%             \\
% -- Code Transformations        & ---                     & ---                \\
% -- RLFixer Enhancements        & ---                     & ---                \\
% -- Field Overwrite Handling    & ---                     & ---                \\
% \bottomrule
% \end{tabular}
% \end{figure}
\begin{figure}[t]
  \centering\small
  \resizebox{\columnwidth}{!}{%
  \begin{tabular}{l|c|c|c|c|c|c}
    \toprule
    \multirow{2}{*}{\lower0.5\normalbaselineskip\hbox{\textbf{Configuration}}}
      & \multicolumn{3}{c|}{\textbf{Leak warnings}}
      & \multicolumn{2}{c|}{\textbf{Fixed warnings}}
      & \multirow{2}{*}{\mbox{\begin{varwidth}{1cm}\textbf{Repair rate}\end{varwidth}}} \\ 
    \cmidrule(lr){2-4}\cmidrule(lr){5-6}
      & \textbf{CL} & \textbf{XE} & \(\mathbf{XR}\)
      & \(\mathbf{F_{CL}}\) & \(\mathbf{F_{XE}}\)
      &  \\ 
    \midrule

    % ── \OurTool ──
    \multirow{2}{*}{\lower0.5\normalbaselineskip\hbox{\OurTool}}
      & \multicolumn{3}{c|}{\BestTot}
      & \multicolumn{2}{c|}{\BestFixTot}
      & \multirow{2}{*}{\BestR\%} \\ 
    \cmidrule(lr){2-4}\cmidrule(lr){5-6}
      & \BestCL & \BestXE & \BestXR
      & \BestFixCL & \BestFixXE
      & \\ 
    \midrule

    % ── Code Transformations ──
    \multirow{2}{*}{\lower0.5\normalbaselineskip\hbox{$-$ Code Transformations}}
      & \multicolumn{3}{c|}{\AOneTot}
      & \multicolumn{2}{c|}{\AOneFixTot}
      & \multirow{2}{*}{\AOneR\%} \\ 
    \cmidrule(lr){2-4}\cmidrule(lr){5-6}
      & \AOneCL & \AOneXE & \AOneXR
      & \AOneFixCL & \AOneFixXE
      & \\ 
    \midrule

    % ── RLFixer Enhancements ──
    \multirow{2}{*}{\lower0.5\normalbaselineskip\hbox{$-$ RLFixer Enhancements}}
      & \multicolumn{3}{c|}{\ATwoTot}
      & \multicolumn{2}{c|}{\ATwoFixTot}
      & \multirow{2}{*}{\ATwoR\%} \\ 
    \cmidrule(lr){2-4}\cmidrule(lr){5-6}
      & \ATwoCL & \ATwoXE & \ATwoXR
      & \ATwoFixCL & \ATwoFixXE
      & \\ 
    \midrule

    % ── Field Overwrite Handling ──
    \multirow{2}{*}{\lower0.5\normalbaselineskip\hbox{$-$ Field Overwrite Handling}}
      & \multicolumn{3}{c|}{\AThreeTot}
      & \multicolumn{2}{c|}{\AThreeFixTot}
      & \multirow{2}{*}{\AThreeR\%} \\ 
    \cmidrule(lr){2-4}\cmidrule(lr){5-6}
      & \AThreeCL & \AThreeXE & \AThreeXR
      & \AThreeFixCL & \AThreeFixXE
      & \\ 
    \bottomrule
  \end{tabular}
  }
  \caption{Ablation study: impact of disabling components of \ourTool. ``Fixed warnings''
    values are \emph{weighted fix counts} (\cref{sec:weighted-fix-count}).}
\mde{When combined, are the rows of this table all the differences between
  \ourTool and \RLCIRLCRLFixer?  Answer the question explicitly.  If it
  isn't, why not?}\mde{Is ``field overwrite
  handling'' equivalent to ``RLC improvements''?  If so, note that fact.}
  \label{tab:ablation-results}
\end{figure}

% \begin{figure}[ht]
%   \centering\small
%   \caption{Impact of shifting leaks across configurations.\todo{I don't intend to keep this table, but just for insider insights. The take-way is we actually fix a lot of shifted leaks, but after mapping them to root library leaks, the number becomes small (125). And even though we are fixing many leaks on the wrappers the overall impact in terms of weighted fixed leaks is not that high.}}
%   \label{tab:shifted-impact}
%   \resizebox{\columnwidth}{!}{%
%     \begin{tabular}{lcccc}
%       \toprule
%       \textbf{Config.} & \textbf{Shifted Raw} & \textbf{Fixed Shifted Raw} & \textbf{Mapped Shifted} & \textbf{Fixed Mapped Shifted} \\
%       \midrule
%       \RLCIRLCRLFixer & 543 & 28 & 100 & 7 \\
%       \OurTool & 838 & 414 & 125 & 53 \\
%       AB1 -- Code Transformation & 543 & 152 & 100 & 34 \\
%       AB2 -- RLFixer Enhancements & 838 & 46 & 125 & 15 \\
%       AB3 -- Field Overwrite Handling & 838 & 414 & 125 & 53 \\
%       \bottomrule
%     \end{tabular}%
%   }
% \end{figure}

% LocalWords:  lr

To measure the contribution of individual components of \ourTool (RQ3), we conduct an ablation study using \ourTool as a base and
selectively disabling key modules:
\begin{enumerate}[]
    \item \textbf{Code Transformations} (\cref{sec:implementation-code-transforms}),
    \item \textbf{RLFixer Enhancements} (\cref{sec:implementation-rlfixer-enhancements}),
    \item \textbf{Field Overwrite Handling} (\cref{sec:implementation-field-overwrite}).
\end{enumerate}
The results are shown in \Cref{tab:ablation-results}.  The numbers are non-trivial to interpret, as disabling certain features may change the total number and location of warnings reported (see \cref{sec:weighted-fix-count}), which has downstream impacts on repair effectiveness.  We explain the results for each configuration below.

When code transformations were disabled, we saw the same number of wrapper types discovered as in \RLCIRLCRLFixer, as finalizer methods are not inserted.  This led to 81 fewer resolved warnings on library resources (they instead appear as core leaks), decreasing the resolution rate to \AOneR\%.  The raw number of core wrapper leak warnings decreases from 838 in \ourTool to 543 with code transformations disabled, with only 152 of them being repaired instead of 412.  The magnitude of this improvement is understated in \Cref{tab:ablation-results} due to use of weighted fix counts (previously discussed in \cref{sec:eval-results}).

When disabling RLFixer enhancements, repair effectiveness is significantly decreased (from 1025 fixed leaks to 814), reducing the resolution rate to \ATwoR\%.  Finally, with field overwrite handling disabled, we see an increase of 81 in \textbf{XE} leaks, due to the lack of filtering of false positives from constructors.  And, there is a decrease in repaired \textbf{XE} leaks (from 62 to 5), due to absence of our new repair pattern for field overwrites, leading to an overall reduction of the resolution rate to \AThreeR\%.

Overall, we see that all components of \ourTool contribute significantly to its effectiveness.

\revised{
\subsection{Run time}
\label{sec:runtime}
We run our experiments on an Ubuntu 20.04.6 LTS cloud VM with 16 vCPUs and 60 GB RAM\@.
Across our evaluation set, the end-to-end \ourTool pipeline averages \textbf{582} seconds per project (\cref{tab:arodnap-timings}).
On average, the second \textsc{RLCI}+\textsc{RLC} pass takes less time than the initial/final RLCI+RLC run because we execute this second pass only on projects where a code transformation was applied.

\begin{figure}[t]
  \centering\small
  % \resizebox{\columnwidth}{!}{%
  \begin{tabular*}{.8\columnwidth}{@{\extracolsep{\fill}} l r @{}}
    \toprule
    \textbf{Stage} & \textbf{Mean time (s)} \\
    \midrule
    RLCI+RLC (first pass) & 211 \\
    Code transformations                     & 15  \\
    RLCI+RLC (second pass) & 78  \\
    Enhanced RLFixer                  & 52  \\
    Patch validation  & 226 \\
    \midrule
    \textbf{Total (per project)}             & \textbf{582} \\
    \bottomrule
  \end{tabular*}
  % }
  \caption{Breakdown of per-project average run time.}
  \label{tab:arodnap-timings}
% \mde{Validation also includes running the tests.  That needs to be added,
%   maybe as a new row.}
\end{figure}

% LocalWords:  RLCI

}

\subsection{Patch Validation}\label{sec:patch-validation-experiment}

% \mde{This section is called ``repair validation'', but
%   \cref{fig:overview-pipeline} uses ``patch validation''.  Make them
%   consistent, probably using ``repair validation''.}

As discussed in \cref{sec:patch-materialization}, \ourTool automatically generates code patches for each repair and 
\revised{validates them using the protocol in \cref{sec:patch-validation-description}.
In contrast, prior work~\cite{UttureP2023} validated only a subset of RLFixer-generated hints.  In 59 cases
% \todo{Sanjay: provide exact number. SM: 49 patch materialization failures, 10 merge conflicts}
with especially complex code structures (49 deeply nested control flow; 10 patch conflicts), automatic materialization did not yield a compilable patch;
for those, we applied the RLFixer hints manually, following the hint template.
\manu{I wrote this para for the major revision deadline; please check}After automatic materialization, we ran static and dynamic validation for all patches generated by \OurTool with all functionality enabled, and found that over 99\% of the patches were validated.
% (see below).  We did not run validation for all of our baseline and ablation configurations due to the manual work required to fix the patch materialization outputs for each configuration.

\mde{Reframe this section.  Do \emph{not} say that \ourTool is sometimes
  incorrect.  Rather, here give data about how often the ``Repair validation''
  step of \cref{fig:overview-pipeline} actually did anything.  Any effect
  of the validation should already be reported in the numbers earlier in
  the paper.}

\textbf{Static validation.} For \ourTool, the generated patches eliminate the reported leak in \textbf{99.1\%} of cases (\(1016/1025\)).
Patch materialization failed in \(9/1025\) attempts, chiefly due to finalizer visibility (private \<close()> methods)
and cases where the repair template structure was insufficient.

\textbf{Dynamic validation.} Across 285 projects, \(11{,}929\) previously passing JUnit tests were re-run; \(7\) newly failed post-repair (an \(\approx 0.06\%\) failure rate).  Five failures were due to the field-to-local conversion, as those fields were accessed elsewhere via reflection.  The other two cases stemmed from control-flow-dependent resource acquisition and finalizer injection closing resources in the wrong order.\manu{For the non-reflection cases, where the failures already present in RLFixer?  May be good to say that \OurTool did not introduce those failure scenarios. SM: The injection of finalizer is from \OurTool, so that failure is new. Previously, that leak was reported on the field assignment, so RLFixer would not have been able to fix it. And for the other failure, as they validated the patches manually, this complex conditional resource allocation were left to developer to handle.}
As a caveat, the overall code coverage of these tests is low (12.9\% statement coverage), and hence only 11\% of Arodnap's patches (107/1,014) were executed by the tests.

% \todo{What is going on in this last category, Sanjay?  Is it the LLM that is wrong or our generated patch hint?}.
In short, we successfully validated nearly all repairs generated by \ourTool; static / dynamic validation and code review should be performed before such repairs are merged.}

\revised{
\subsection{Example Fixed Leaks}\label{sec:case-fixed}
As in the motivating example (\cref{fig:mot-example}),
\ourTool's transformations make ownership explicit and can \emph{shift} warnings to more actionable sites
(\cref{sec:weighted-fix-count}). We highlight two additional patterns taken from the NJR dataset that prior logic did not repair
but \ourTool now fixes.

\subsubsection{Safe Pre-close Before Field Overwrite}
\label{sec:case-preclose}
Reassigning a resource field can leak the prior value. When the field is private, every write stores a freshly
allocated resource, and no aliases can escape, \ourTool inserts a conditional cleanup \emph{before} the write (\cref{lst:case-preclose-tsocket}).
}
\begin{figure}
\begin{lstlisting}[language=diff]
 class AbstractParserTables {
   private Writer f = null;

   String toSourceFile(String fileName) {
  	 File file = new File(fileName);
+    if (f != null) {
+        f.close();
+    }
     f = new BufferedWriter(new FileWriter(file));
     // ...
   }
 }
\end{lstlisting}
\caption{Inserting a pre-close before overwriting a resource field.
Adapted from benchmark \<url270fc4f5ee\_ykcilborw\_Joust\_tgz>, file \<AbstractParserTables.java>.}
% \mde{The paper is inconsistent about what pre-close looks like.  Here there
  % is a \<try>--\<catch>, but in \cref{lst:mot-example-fixed} there is not.}
\label{lst:case-preclose-tsocket}
\end{figure}
\revised{
  \newline
\noindent\textit{Why this was hard before.} Without containment and ownership information,
a pre-close risks use-after-close via surviving aliases, so prior repair avoids inserting it.
% Our guards discharge that risk, and the patch is validated by compilation, an RLC re-run (no new warnings),
% and unchanged JUnit outcomes.

\subsubsection{Containment Proves an Accessor, Enabling Client-Side Repair}
\label{sec:case-containment}
~As shown in \cref{lst:case-containment-user}, the user-defined class \<Task> caches an incoming resource in a private field, never lets it escape, and exposes no finalizer.
Treating such a class as an \emph{owner} blocks repair; with containment, we classify it as an \emph{accessor} and fix at the client \<startPuppeteer>.
}
\begin{figure}
\begin{lstlisting}[language=diff]
 class Task extends TimerTask {
   private final Puppeteer m_Puppeteer;
   public Task(Puppeteer puppeteer) { m_Puppeteer = puppeteer; }
   public void run() { // uses m_Puppeteer};
 }
 
 class ActorsTest {
   void startPuppeteer() {
-    Puppeteer puppeteer = new Puppeteer("localhost");
-    (new java.util.Timer("Puppeteer")).schedule(new Task(puppeteer), 0, 1000);
+    Puppeteer puppeteer = null;
+    try {
+      puppeteer = new Puppeteer("localhost");
+      (new java.util.Timer("Puppeteer")).schedule(new Task(puppeteer), 0, 1000);
+    } finally {
+      if (puppeteer != null) puppeteer.finish();
+    }
   }
 }
\end{lstlisting}
\caption{Client-side try/finally for a resource accessor that does not own the
  resource.
  Adapted from benchmark \<urlc98c3b97d2\_Trimax\_venta\_tgz>, file \<ActorsTest.java>.}
\label{lst:case-containment-user}
\end{figure}
\revised{
  \newline
\noindent\textit{Why this was hard before.} Without containment, the wrapper is conservatively treated as a potential owner
or rejected for lacking a finalizer, so no safe fix is emitted. Containment shows the field does not escape; the client-side
\<try/finally> is then sound under our ownership model.
%  and passes our static and dynamic validation.
}
\subsection{Remaining Unfixed Leaks: Case Study}

To address RQ4 and better understand \ourTool's limitations, we manually
inspected 100 randomly chosen locations of resource leak warnings that
remained unfixed by \ourTool.

63\% of the unrepaired cases would require more global analysis and/or transformation to repair.
In 49 cases, we found that the resource
truly escaped the local scope of the warning to some longer-lived object or
data structure.  Repairing such cases could require significant changes
across the codebase and advances in verification reasoning to prove safety.
In 14 other cases, there was an overwrite of an owning field where the
checks of \cref{sec:implementation-field-overwrite} could not prove it was
safe to close the field before the overwrite.  In most of these cases, the
field was not private, or a resource stored in the field was passed in from outside the enclosing
class, so more global analysis would be required to prove safety of the
repair.

% \todo{Adding a transformation that makes a field
%   private would further improve the experimental results; we should do it
%   while waiting for the reviews.}

The remaining 37\% of unrepaired cases could be addressed with further
engineering improvements to RLFixer that are orthogonal to our
contribution.
In 28 cases, the repair could not be performed due to the need for more complex repair templates in RLFixer, e.g., to handle resources allocated in loops or exceptions thrown from a constructor after a field assignment.  Finally, the remaining 9 cases could not be repaired due to remaining limitations in RLFixer's logic to match leak warnings to WALA IR instructions.

\section{Limitations and Threats to Validity}
\label{sec:limitations}

\subsubsection{Limitations}
Currently, \ourTool cannot repair leak warnings related to creation of
fresh obligations on wrappers with non-final \Owning fields.
(RLC expresses this with a \CreatesMustCallFor
annotation~\cite{KelloggSSE2021}.)
An overwrite of such a field could ``reset'' the obligation on a wrapper
after its finalizer method has already been called, necessitating another
finalizer call.  RLC is currently very imprecise in reasoning about such
cases, and hence the reports are not amenable to repair; Shadab et al.'s
work on inference also ignored such warnings~\cite{ShadabGTEKLLS2023}.

\revised{
\ourTool's transformations are sound and conservative, and therefore they
miss some opportunities.
Invoking the injected finalizer can interact with framework lifecycles;
we therefore inject it only when there is a relevant RLC warning.
For pre-close on reassigned fields, we apply
the edit only when the field is \emph{private}, every write stores a freshly allocated resource, and field containment (\cref{sec:case-containment}) holds; otherwise we
skip to avoid alias-related use-after-close risks.}

\subsubsection{Threats to Validity}
Regarding external validity, our evaluation is conducted on 285 Java 8 benchmarks from the NJR-1 dataset, which was used in prior resource leak repair research~\cite{UttureP2023}. While the dataset is diverse,
our results may not fully generalize to programs targeting more modern Java versions, Android applications, or other programming ecosystems with different resource management idioms.

Regarding internal validity, to compare leak warnings across configurations, we use an automated mapping process that matches leaks on wrapper types to the corresponding leaking library resources (\cref{sec:weighted-fix-count}).  When the automation failed to establish a match, we manually completed the mapping.  There may still be minor inconsistencies in our mapping, due to imperfect alias resolution or complex control flow.

Beyond the correctness of \ourTool's implementation, our results rely on the correctness of RLC, RLCI, RLFixer, and supporting tools like JavaParser and WALA\@. Limitations or bugs in any of these components may affect the accuracy of detection or the applicability of repairs.  We have validated all of our generated repairs against RLC (\cref{sec:patch-validation-experiment}), a strong consistency check for the full \ourTool pipeline.
% \todo{Why not ``all''? ``Nearly'' is a weasel word that will annoy reviewers. SM: All patches are validated by RLC.}
% Finally, while LLM-generated patches are validated for syntactic correctness and
% successful re-analysis, we do not measure their readability or acceptability from a developer's perspective.

% LocalWords:  NJR RLCI WALA

\section{Related Work}
\label{sec:relatedwork}
The research most related to our work spans \emph{static analysis for
defect detection}, \emph{specification inference}, and \emph{automated program
repair (APR)}. We discuss these areas in turn.

\textbf{Static Analysis for Detection.} Static analysis techniques have
long been employed to detect resource management errors.
RLFixer~\cite{UttureP2023} was specifically tested with the Infer
\cite{CalcagnoDDGHLOPPR2015}, SpotBugs~\cite{HovemeyerP2004},
CodeGuru~\cite{CodeGuru}, PMD~\cite{PMD}, and RLC \cite{KelloggSSE2021}
leak detectors.  These techniques were built based on earlier research on
static leak detection for Java like that of Torlak and
Chandra~\cite{TorlakC10}.  Other languages have their own leak detection
tools, e.g., the Clang Static Analyzer~\cite{ClangAnalyzer} for C/C++.  Our
work focuses on repair of leaks reported by RLC, as its corresponding
inference technique~\cite{ShadabGTEKLLS2023} uniquely and automatically
exposes key information needed to repair leaks on wrapper types.\looseness=-1

% Facebook's \textsc{Infer}~\cite{CalcagnoSPRINGER2015}, \textsc{SpotBugs}~\cite{AyewahIEEES2008}, Google's \textsc{Error Prone}~\cite{ErrorProne} and the Checker Framework's
% \textsc{Resource Leak Checker (RLC)}~\cite{PapiACPE2008} employ ownership reasoning, path-sensitive flow analysis, and typestate tracking to report unclosed resource instances.
% Speacialised research tools and more precise inter-procedural analyses, such as \textsc{FlowDroid}~\cite{ArztPLDI2014}, \textsc{TaintDroid}~\cite{EnckACMTCS2014} and
% \textsc{PlumbDroid}~\cite{BhattJSS2022} track resources through Android callbacks and lifecycle events, while the \textsc{Clang Static Analyzer}~\cite{ClangAnalyzer} offers an
% analogous analysis for C/C++ projects. Our work targets Java like \textsc{Infer} but boosts recall by integrating automatically inferred annotations (\S\ref{sec:inference})
% and by chaining a repair phase that rewrites resource handling code (\S\ref{sec:rlfixer}).

\textbf{Specification Inference.} The inference technique for RLC~\cite{ShadabGTEKLLS2023} is built on the whole-program inference framework of Kellogg et al.~\cite{KelloggDNAE2023}.  Other recent approaches to annotation inference have been based on black-box search~\cite{DBLP:conf/sigsoft/KarimipourPCS23}, information retrieval~\cite{DBLP:journals/pacmpl/WuL24}, and machine-learning~\cite{HellendoornBBA2018,PengGLGLZL2022,PradelGLC2020}.  \ourTool uses the extant RLC inference without modification, and improved inference techniques could be easily incorporated.  Recent work on mining API-level resource patterns{\,---\,}e.g.\ MiROK's large-scale extraction of acquisition-release
pairs~\cite{WangESEC2023}, MAPO's protocol mining~\cite{ZhongECOOP2009}{\,---\,}provides a complementary source of
cleanup specifications that could be used to extend the applicability of \ourTool in the future.

% Our work leverages these insights by using inference to fill gaps in resource lifecycle obligations and complement the static analysis with
% transformation rules that make such obligations enforceable.

\textbf{Automated Repair Techniques.} Automated repair has been studied across multiple paradigms, and we categorize the most relevant work into static template-based repair and machine
learning-driven repair.

\textit{Static Analysis Template-Based Repair.} Template-driven \textit{static} APR has evolved along two complementary lines. First, \emph{general-purpose} systems such as \textsc{GenProg}~\cite{LeGouesNFW2012}, \textsc{SemFix}~\cite{NguyenICSE2013}, and \textsc{CoCoNuT}~\cite{LutellierISSTA2020} explore very large patch spaces via mutation or semantic search guided by test outcomes.
Because resource leaks seldom manifest as failing tests, these techniques are largely ineffective in our setting~\cite{UttureP2023}. Van Tonder and Le Goues showed that separation-logic proofs can
facilitate synthesis of verified patches for heap anomalies, including leaks, in C programs \cite{vanTonderLG2018}. 
% RLFixer pushes that idea further with alias-aware escape analysis and
% specialised leak-repair templates, albeit with a deliberately conservative scope to safeguard correctness~\cite{UttureP2023}.
Second, a broad ecosystem of \emph{domain-specific static APR tools} shows that carefully crafted templates plus static reasoning can repair defects even when no failing tests exist.  
\textsc{MemFix} formulates C memory-deallocation faults (leaks, double-frees, use-after-frees) as an exact-cover problem over allocation-free pairs and solves it with a SAT solver~\cite{LeeESEC2018}.  
\textsc{ARC} applies a genetic search that mutates Java synchronization constructs to eliminate deadlocks and data races, then prunes excess locks for performance~\cite{KelkMUSEPAT2013}.  
\textsc{NPEFix} dynamically guards or substitutes risky dereferences to avert null-pointer crashes in Java~\cite{LeeICSE2022}.  Our tool \ourTool is a domain-specific approach targeted at extending repair of resource leaks to wrapper types.
% \textsc{Prophet} learns a probabilistic patch model from thousands of human commits and uses it to rank template-generated fixes~\cite{LongPOPL2016}.  
% These successes demonstrate that domain-aware templates, guided by static analysis or learned patch priors, can yield high-quality repairs without test oracles{\,---\,}an insight we generalise to resource-lifecycle bugs.

\textit{LLM-Based Repair Techniques.} Recent work couples static analysis with large-language-model (LLM) patch generation. \textsc{InferFix} augments \textsc{Infer} warnings with retrieval-based
context before feeding them to a fine-tuned LLM, achieving strong results on test-oriented benchmarks~\cite{JinESEC2023}, while \textsc{FixrLeak} deploys a prompt-engineering workflow at Uber that
turns static leak warnings into try-with-resources rewrites for Java services~\cite{ZhangUSP2024}.  Neither of these approaches is applicable to the wrapper type leaks targeted by \ourTool.  Transformer-based systems such as \textsc{TFix}~\cite{BerabiPMLR2021}, \textsc{CURE}~\cite{ZhongICSE2022},
and \textsc{ReCoder}~\cite{ZhuESEC2021} learn edit patterns from historical commits. These models excel at syntactic and localized edits but often lack deep inter-procedural reasoning, limiting their potential effectiveness for wrapper type leaks.  Our approach applies code transformations, inference, and targeted static analysis to discover and repair leaks involving wrapper types across a large program scope.

% their effectiveness
% on latent resource leaks that demand whole-program reasoning. Our hybrid strategy keeps static analysis ``in the loop'' to validate lifecycle constraints, delegating only the syntactic realisation of a fix
% (e.g., complete block replacements or `try'/`catch' scaffolding) to the LLM, thereby balancing soundness and flexibility.

% \textbf{Other Related Approaches.} Escape analysis{\,---\,}originally devised to decide whether an object can be stack-allocated or synchronization elided~\cite{ChoiOOPSLA1999}{\,---\,}forms the conceptual backbone of our
% \emph{resource-escape} reasoning, except that we classify the \emph{targets} of an assignment (fields, parameters, collections) to enforce safe disposal rather than memory safety per se.
% Additionally, \textsc{RESIN}~\cite{ChangOSDI2022} provides a production-scale, dynamic memory-leak detection and automatic mitigation service in the cloud, complementing our static, sourcelevel repair pipeline.

% Our pipeline extends this static-analysis-informed tradition to Java resource leaks by unifying specification inference, template-based transformation, and verification in a fully automated workflow.

\revised{
\textbf{Android Resource Leaks.}
For Android, datasets such as \textsc{DroidLeaks} curate real resource-leak defects and are widely used for evaluation~\cite{LiuESE2019}.
Analyses like \textsc{PlumbDroid} detect and automatically repair Android resource leaks by reasoning over event-driven control flow and lifecycle callbacks~\cite{BhattJSS2022}.
While not leak-specific, \textsc{FixDroid} is an Android Studio assistant that flags security/privacy pitfalls and offers quick fixes~\cite{NguyenCCS2017}.
\ourTool, by contrast, targets general Java projects and wrapper-based ownership patterns.
Its analyses and transformations operate at the language level and are agnostic to Android-specific lifecycles, making these lines of work complementary rather than overlapping.
}

% LocalWords:  CodeGuru MiROK's MAPO's GenProg SemFix CoCoNuT Tonder Goues
% LocalWords:  MemFix NPEFix InferFix FixrLeak TFix ReCoder DroidLeaks
% LocalWords:  PlumbDroid FixDroid

\section{Conclusion}

\OurTool is a technique and tool that extends resource leak repair for Java to apply to resource wrappers.  Java programs often store resources in fields of wrappers, and repairing leaks of such resources requires techniques targeted at the wrapper types and their fields.
\OurTool demonstrates that new static analysis techniques, not just better integration, are needed to close the gap on resource leak repair for wrappers.
Through a combination of code transformations, new repair templates, and enhanced reasoning about fields during both leak detection and repair, \ourTool achieved a leak resolution rate of \BestR\%, improving over the \BOneR\% rate of the prior state of the art.\looseness=-1

\section*{Acknowledgements}

This material is based upon work supported by the Defense Advanced Research
Projects Agency (DARPA) under Agreement No. HR00112590132, the National
Science Foundation under grants CCF-2223826, CCF-2312262, CCF-2312263, and
CNS-2120070, a gift from Oracle Labs, and a Google Research Award.

% \begin{acks}
% \end{acks}

\bibliographystyle{IEEETran}
\bibliography{local,plume-bib/bibstring-abbrev,plume-bib/types,plume-bib/dispatch,plume-bib/ernst,plume-bib/invariants,plume-bib/other,plume-bib/program-analysis,plume-bib/soft-eng,plume-bib/crossrefs}


\end{document}